\documentclass[useAMS, a4, usenatbib]{mn2e}
\input psfig.sty
\usepackage{graphicx}
\usepackage{floatflt, epsfig}

\usepackage{epstopdf}

\usepackage{longtable}
\usepackage{sidecap}
\usepackage{subfigure}
\usepackage{footnote}
\usepackage{amsmath}

\usepackage{tabulary}
\usepackage{multirow}

\def \chisq  {\ifmmode  \chi^2   \else  $\chi^2$  \fi}  
\def \spose#1{\hbox  to 0pt{#1\hss}}  
\def \lta{\mathrel{\spose{\lower 3pt\hbox{$\sim$}}\raise  2.0pt\hbox{$<$}}}
\def \gta{\mathrel{\spose{\lower  3pt\hbox{$\sim$}}\raise 2.0pt\hbox{$>$}}}

\def \kms {\ifmmode  \,\rm km\,s^{-1} \else $\,\rm km\,s^{-1}  $ \fi }
\def \kpc {\ifmmode  {\rm~kpc}  \else ${\rm~kpc}$\fi}  
\def \pc {\ifmmode  {\rm~pc}  \else ${\rm~pc}$ \fi  }  
\def \Gyr {\ifmmode  {\rm~Gyr}  \else ${\rm~Gyr}$\fi}
\def \Msun {\ifmmode M_{\odot} \else $M_{\odot}$ \fi} 
\def \Lsun {\ifmmode L_{\odot} \else $L_{\odot}$ \fi} 
\def \Rsun {\ifmmode R_{\odot} \else $R_{\odot}$ \fi} 
\def \Msunpyr {\ifmmode M_{\odot}{\rm~yr}^{-1} \else $M_{\odot}{\rm~yr}^{-1}$ \fi} 
\def \hMsun {\ifmmode h^{-1}\,\rm M_{\odot} \else $h^{-1}\,\rm M_{\odot}$ \fi}

\def \LCDM {\ifmmode \Lambda{\rm CDM} \else $\Lambda{\rm CDM}$ \fi}
\def \sig8 {\ifmmode \sigma_8 \else $\sigma_8$ \fi} 
\def \OmegaM {\ifmmode \Omega_{\rm M} \else $\Omega_{\rm M}$ \fi} 
\def \OmegaL {\ifmmode \Omega_{\rm \Lambda} \else $\Omega_{\rm \Lambda}$\fi} 
\def \Deltavir {\ifmmode \Delta_{\rm vir} \else $\Delta_{\rm vir}$ \fi}
\def \rhocrit {\ifmmode \rho_{\rm crit} \else $\rho_{\rm crit}$ \fi}
\def \rhou {\ifmmode \rho_{\rm u} \else $\rho_{\rm u}$ \fi}
\def \zc {\ifmmode z_{\rm c} \else $z_{\rm c}$ \fi}

\def \rhos {\ifmmode \rho_{\rm s} \else $\rho_{\rm s}$ \fi} 
\def \rs {\ifmmode r_{\rm s} \else $r_{\rm s}$ \fi} 
\def \cvir {\ifmmode c_{\rm vir} \else $c_{\rm vir}$ \fi} 
\def \Rvir {\ifmmode r_{\rm vir} \else $R_{\rm vir}$ \fi}
\def \Vvir {\ifmmode V_{\rm  vir} \else  $V_{\rm vir}$  \fi} 
\def \Mvir {\ifmmode M_{\rm  vir} \else $M_{\rm  vir}$ \fi}  
\def \Nvir {\ifmmode N_{\rm  vir} \else $N_{\rm  vir}$ \fi}  
\def \Jvir {\ifmmode J_{\rm vir} \else $J_{\rm vir}$ \fi} 
\def \Evir {\ifmmode E_{\rm vir} \else $E_{\rm vir}$ \fi} 
\def \vvir {\ifmmode v_{\rm vir} \else $v_{\rm vir}$ \fi} 
\def \lam {\ifmmode \lambda  \else $\lambda$ \fi} 
\def \lamp {\ifmmode \lambda^{\prime} \else $\lambda^{\prime}$  \fi} 
\def \Vmax {\ifmmode V_{\rm  max} \else  $V_{\rm max}$  \fi} 
\def \Mdm {\ifmmode M_{\rm  dm} \else $M_{\rm  dm}$ \fi}

\def \Mgas {\ifmmode M_{\rm gas} \else $M_{\rm gas}$ \fi} 
\def \Mcg {\ifmmode M_{\rm cg} \else $M_{\rm cg}$\fi} 
\def \Mhg {\ifmmode M_{\rm hg} \else $M_{\rm hg}$ \fi} 
\def \Mdisc {\ifmmode M_{\rm disc} \else $M_{\rm disc}$ \fi} 
\def \Md {\ifmmode M_{\rm d} \else $M_{\rm d}$ \fi} 
\def \Mda {\ifmmode M_{\rm d,0\%} \else $M_{\rm d,0\%}$ \fi} 
\def \Mdb {\ifmmode M_{\rm d,20\%} \else $M_{\rm d,20\%}$ \fi} 
\def \Mdc {\ifmmode M_{\rm d,40\%} \else $M_{\rm d,40\%}$ \fi} 
\def \md {\ifmmode m_{\rm d} \else $m_{\rm d}$ \fi} 
\def \Mb {\ifmmode M_{\rm b} \else $M_{\rm b}$ \fi} 
\def \Mbh {\ifmmode M_{\rm b,pri} \else $M_{\rm b,pri}$ \fi} 
\def \Mbs {\ifmmode M_{\rm b,sat} \else $M_{\rm b,sat}$ \fi} 
\def \zo {\ifmmode z_{0} \else $z_{0}$ \fi} 
\def \rd {\ifmmode r_{\rm d} \else $r_{\rm d}$ \fi}
\def \rg {\ifmmode r_{\rm g} \else $r_{\rm g}$ \fi}
\def \rb {\ifmmode r_{\rm b} \else $r_{\rm b}$\fi}
\def \rs {\ifmmode r_{\rm s} \else $r_{\rm s}$\fi}
\def \rc {\ifmmode r_{\rm c} \else $r_{\rm c}$\fi}
\def \rvir {\ifmmode r_{\rm vir} \else $r_{\rm vir}$\fi}
\def \rbh {\ifmmode r_{\rm b,pri} \else $r_{\rm b,pri}$ \fi} 
\def \rbs {\ifmmode r_{\rm b,sat} \else $r_{\rm b,sat}$ \fi}

\title[From Discs to Bulges] 
{From Discs to Bulges: effect of mergers on the morphology of galaxies}

\author[Kannan et al.] {Rahul  Kannan$^{1,2}$\thanks{kannanr@mit.edu}, Andrea V. Macci\`o$^{1}$\thanks{maccio@mpia.de}, Fabio Fontanot$^{3,4}$, 
\newauthor{Benjamin P. Moster$^5$, 
Wouter Karman$^{6}$, Rachel S. Somerville$^{7}$}\\
  $^1$ Max-Planck-Institut f\"ur Astronomie, K\"onigstuhl 17, 69117 Heidelberg, Germany\\ 
  $^2$ Department of Physics, Kavli Institute for Astrophysics $\&$ Space Research, Massachusetts Institute of Technology, Cambridge, MA 02139, USA \\
  $^3$ INAF - Astronomical Observatory of Trieste, via G.B. Tiepolo 11, I-34143 Trieste, Italy \\
  $^4$ HITS - Heidelberger Institut f\"ur Theoretische Studien, Schloss-Wolfsbrunnenweg 35, 69118 Heidelberg, Germany \\
  $^5$ Kavli Institute for Cosmology, Institute of Astronomy, University of Cambridge, Madingley Road, Cambridge CB3 0HA, UK\\
  $^6$ Kapteyn Institute, University of Groningen, P.O. Box 800, 9700 AV Groningen, The Netherlands\\
  $^7$ Department of Physics and Astronomy, Rutgers University, 136 Frelinghuysen Rd., Piscataway, NJ 08854, USA
\\
}

\begin{document} 
              
\date{\today}
              
\pagerange{\pageref{firstpage}--\pageref{lastpage}}\pubyear{2015} 
 
\maketitle 

\label{firstpage}
             
\begin{abstract}

We study the effect of mergers on the morphology of
galaxies by means of the simulated merger tree approach first
proposed by Moster et al. This method combines N-body cosmological
simulations and semi-analytic techniques to extract realistic initial conditions for galaxy mergers. These are
then evolved using high resolution hydrodynamical
simulations, which include dark matter, stars, cold gas in the disc and hot gas in the halo. 
We show that the satellite mass accretion is not as effective as previously thought, as there is substantial stellar stripping before the final merger. 
The fraction of stellar disc mass transferred
to the bulge is quite low, even in the case of a major merger, mainly due to the dispersion of part of the stellar disc mass into the
halo. We confirm the findings of Hopkins et al., that a gas rich disc is able to survive major mergers more efficiently.
The enhanced star formation associated
with the merger is not localised to the bulge of galaxy, but
a substantial fraction takes place in the disc too.
The inclusion of the hot gas reservoir in the galaxy model 
contributes to reducing the efficiency of bulge
formation.
Overall, our findings suggest that mergers are not as efficient as previously
thought in transforming discs into bulges. This possibly alleviates
some of the tensions between observations of bulgeless galaxies and the
hierarchical scenario for structure formation.

\end{abstract}


\setcounter{footnote}{1}

\section{Introduction}
\label{sec:intro}

Galaxy  morphologies  constitute  one  of  the  earliest  attempts  to
classify galaxies, according  to the relative prominence  of their two
main  components, i.e.  the (spheroidal)  bulge and  the (exponential)
disc.     This      scheme     was     originally      proposed     by
\citet{1926ApJ....64..321H}.   The  nature  of the  link  between  the
morphological properties  of a  galaxy and its  cosmologically defined
evolutionary track  is a fundamental  question in the field  of galaxy
formation and evolution.

The standard model of galaxy formation assumes that the gas in
  dark matter  potential wells  cools to  the center  of the  well and
  forms a disc, out of which stars form.
Violent  dissipative processes  such as  mergers and  close encounters
remove the angular momentum from the disc fuelling bulge formation. In
a $\Lambda$ cold dark matter universe ($\Lambda$CDM), dark matter (DM)
structures grow  hierarchically \citep{1978MNRAS.183..341W},  with the
smaller dark  matter haloes  forming first and  later merging  to form
bigger ones:  this makes  interactions between galaxies  a fundamental
and inescapable  process of  galaxy evolution. Mergers  are considered
the  origin  of  the  so   called  ``classical''  bulges  (i.e.  whose
properties   are    similar   to   Elliptical   galaxies,    see   e.g
\citealt{1983ApJ...266..516D};  \citealt{2013MNRAS.428.3183D}),  while
``pseudo''-bulges  (i.e.  those  characterized by  disc-like  profiles
and/or   kinematics,  see   e.g.  \citealt{2004ARA&A..42..603K})   are
connected with ``in-situ"  processes like gravitational instabilities,
which  leads  to  the  rearrangement  of  the  disc  material  into  a
spheroidal-like structure. Additionally,  a bulge-dominated galaxy may
regrow a  new stellar disc,  if there is  a sufficient supply  of cold
gas, e.g., from the reservoir  of hot gas present in quasi-hydrostatic
equilibrium within the gravitational potential of the dark matter halo
(see  e.g.  \citealt{2011MNRAS.414.1439D})   or  through  cosmological
accretion.  The relative  efficiency  of these  processes dictate  the
morphology of the galaxy.

Many authors  have tried  to figure out  the relative  contribution of
bulge  and  disc   components  of  galaxies  in   the  local  universe
\citep{2009ApJ...696..411W,
  2011ApJ...733L..47F}.   \cite{2009MNRAS.393.1531G}  calculated   the
stellar mass content  and distribution for each  galaxy component, and
showed that,  for galaxies  more massive than  $10^{10}$ \Msun  in the
local universe, 32 per cent of  the total stellar mass is contained in
ellipticals, and the  corresponding values for discs,  bulges and bars
are $36$, $28$ and $4\%$  respectively. Classical bulges contain $25\%$
of the total stellar mass, while pseudo-bulges contain $3\%$ per cent.

\cite{2010ApJ...723...54K}  looked at galaxies in the local neighborhood ($<8$ Mpc) and find four  galaxies consistent  with being
pure  disc galaxies  and 7  galaxies,  including the  Milky Way,  having
pseudo bulges. Given  the dearth of classical bulges  in their sample,
they thus  estimate that around  $58-74 \%$  of the galaxies  in their
sample did  not undergo violent  mergers in  their past and  thus they
claim  that   the  formation  of  these   massive  bulgeless  galaxies
represents  a challenge  for  current models  of  galaxy formation.  A
recent study  of the  morphological mix  in the  SDSS volume  has been
discussed in \cite{2012ApJ...746..160W}; they  provide the fraction of
galaxies showing a given  morphological and activity classification as
a  function of  stellar and  parent halo  masses. They  find that  the
fraction of elliptical galaxies is  a strong function of stellar mass;
it  is also  a strong  function  of halo  mass, but  only for  central
galaxies. This is treated as  evidence for a scenario where elliptical
galaxies are always formed, probably  via mergers, as central galaxies
within  their  halos,  with  satellite  ellipticals  being  previously
central galaxies accreted onto a larger halo.

Different  theoretical  tools  have   been  employed  to  explain  and
understand this  observational evidence. Bulge formation  processes in
Semi Analytic  Models (SAMs)  and their  relative importance  has been
studied    in    detail    in    a    number    of    recent    papers
(\citealt{2010MNRAS.406.1533D};         \citealt{2011MNRAS.416..409F};
\citealt{2013MNRAS.433.2986W};   \citealt{2014MNRAS.444..942P}).   The
general consensus  in these works  is that processes like  mergers and
disc  instabilities  are  key   to  understanding  the  morphology  of
intermediate mass galaxies ($10^{10}<M_{\*}/\Msun<10^{11}$): therefore
our limited understanding of these  mechanisms is a limitation for the
models' ability  of reproducing the morphological  mix. In particular,
\citet{2011MNRAS.416..409F}  showed  that  the observed  abundance  of
massive  galaxies without  a classical  bulge is  consistent with  the
predicted abundance of bulgeless galaxies  only for a model where disc
instability     process     is     not    considered     (see     also
\citealt{2014MNRAS.444..942P}).

If the angular  momentum of a primordial halo is  conserved during its
collapse  then  it  is  sufficient   to  produce  large  discs  (e.g.,
\citealt{1980MNRAS.193..189F};         \citealt{1998MNRAS.295..319M}).
Cooling in  the highly dense  inner regions and dynamical  friction of
orbital satellites dissipates  the angular momentum of  the gas (e.g.,
\citealt{2006MNRAS.372.1525D}),  thus resulting  in  compact discs  in
hydrodynamical N-body  simulations, in which the  rotation curve peaks
at a  few \kpc, in  contradiction to  the rotation curves  of observed
galaxies             (e.g.,             \citealt{1999ApJ...513..555S};
\citealt{2008ASL.....1....7M}   ).  \cite{2005ApJ...622L...9S}
  and  \cite{2006ApJ...645..986R}  showed  that  in  idealized  merger
  simulations with  strong stellar  feedback it is  possible to  get a
  disc   dominated    remnant,   if   the   initial    disc   is   gas
  rich. Furthermore,  recent cosmological  simulations have  also been
  able   to  form   disc  galaxies   using  strong   stellar  feedback
  prescriptions    \citep{2007MNRAS.374.1479G,    2011ApJ...742...76G,
    2012MNRAS.424.1275B,   2012ApJ...756...26M,   2013MNRAS.428..129S,
    2014MNRAS.437.2882K, 2014MNRAS.437.1750M, 2014MNRAS.440L..51C}.

\citet[hereafter  H09]{2009ApJ...691.1168H} quantified  the dependence
of bulge formation on the gas disc fraction and presented a simple toy
model to  account for it.  They showed that  it is possible  to obtain
disc dominated remnants  even for $1:1$ gas rich  mergers. This result
has  major implications  for the  amount of  bulge dominated  galaxies
found in the local universe (\citealt{2009MNRAS.397..802H}).

In this  paper we re-examine  the disc to bulge  transformation during
mergers.   We   adopt   a    hybrid   method,   first   developed   by
\cite{2014MNRAS.437.1027M}   which  is   based   on  high   resolution
hydrodynamical  simulations of  merger systems.  The dark  matter halo
properties and  their orbital  parameters are directly  extracted from
cosmological simulations, while the  properties of the galaxies hosted
by those  halos are  predicted using  a Semi  Analytic Model  (SAM) of
galaxy formation.
 
In this  way we gain  the advantages  of the merger  simulations (high
resolution  and  correct  treatment  of   gas  physics)  and  the  SAM
(cosmological background).  Simultaneously, the computational  cost is
comparably low, so that a meaningful sample can be modelled in a short
amount of time.
Given the typical resolution of these hydrodynamical simulations, this
method is  well suited to resolve  the small scales, relevant  for the
study of the evolution of the stellar components of galaxies and their
scale parameters,  such as the disc  scale length and height,  and for
the  evolution  of galaxy  morphology.   This  approach allows  us  to
achieve the  best resolution  possible within  a reasonable  amount of
time, while  being able to model  a sample of galaxies  in the correct
cosmological context.

We describe  the numerical techniques  used to simulate mergers  in \S
\ref{sec:methods} and also  give a brief introduction of  the SAM used
in this study.  We enumerate the results in \S  \ref{sec:sims} and the
conclusions and discussions are given in \S \ref{sec:discussion}.

\section{Models}
\label{sec:methods}

In this section we briefly describe the methods that have been used in
this paper.  These are,  the simulation code  {\sc pinochio}  that was
employed to generate cosmological merger  trees, the SAM {\sc morgana}
used to  populate the merger trees  with galaxies, the code  to create
initial conditions and  the hydrodynamics code {\sc  gadget-2} used to
perform  the  merger  simulations.  Throughout  this  paper  we  adopt
cosmological  parameters chosen  to  match results  from {\sc  WMAP}-3
\citep{2007ApJS..170..377S}  for  a   flat  $\Lambda$CDM  cosmological
model:            $\Omega_m=0.26$,            $\Omega_{\Lambda}=0.74$,
$h=H_0/(100$~km~s$^{-1}$~Mpc$^{-1})=0.72$,     $\sigma_8=0.77$     and
$n=0.95$.  We adopt a  \citet{2001MNRAS.322..231K} IMF and compute all
stellar masses accordingly.

\subsection{Merger Tree generation : PINOCCHIO}
\label{sec:ncode}

To construct the Dark Matter halo merger trees we make use of the
  pinocchio code  \citep{2002MNRAS.331..587M}. {\sc pinocchio}  uses a
  scheme  based on  Lagrangian perturbation  theory, which  we briefly
  describe  in the  following (see  also \citealt{2002MNRAS.333..623T}
  for a more detailed discussion about the definition of DM haloes and
  merger trees).  A Gaussian  linear density  contrast field  (for the
  desired cosmological  background) is defined  on a cubic  grid, then
  smoothed repeatedly with Gaussian filters.  For each particle on the
  grid,  the  six  nearest  particles are  considered  its  Lagrangian
  neighbours. The collapse time of particles (i.e. the time they first
  belong to a high density,  multi-stream region) is then computed for
  each  point of  the Lagrangian  space, following  the definition  of
  orbit  crossing proposed  in \citet{2002ApJ...564....8M}.  Collapsed
  particles accrete  onto individual DM haloes  or filaments according
  to  their distance  from the  centre of  mass of  neighboring haloes
  (i.e.  those containing  at least  one  of its  neighbours): if  the
  distance  is  smaller  than  a  given fraction  of  halo  size,  the
  collapsed  particle  then  became   an  accreting  particle  of  the
  halo. Filament  particles can be accreted  at later times if  any of
  their  neighbour  became an  accreting  particle.  Moreover, two  DM
  haloes  merge  following  a  similar  prescription,  i.e.  if  their
  distance in Lagrangian space is smaller  than a fraction of the size
  of  the  more  massive  object. Overall  {\sc  pinocchio}  allows  a
  detailed  reconstruction of  the  DM haloes,  with known  positions,
  velocities  and  angular momenta,  and  of  their merger  trees,  in
  excellent  agreement with  the results  of N-body  simulations, (see
  e.g. \citealt{2007ApJ...665..187L}), with a  very fine time sampling
  that provides  tracking of  merging times  without restriction  to a
  fixed grid in time (as  in N-body simulations). However, at variance
  to N-body trees  pinocchio DM haloes are not allowed  to decrease in
  mass, and the code does not  track the evolution of DM substructures
  once  they have  been  accreted  by the  main  halo: however,  these
  differences do  not constitute a  limitation in our case,  since the
  mass  evolution of  DM substructures  is explicitly  tracked by  the
  hydrodynamical simulation.

\subsection{Semi-analytic model: {\sc MORGANA}}
\label{subsec:sam}

The {\sc pinocchio} merger trees have  then been used as input for the
Semi       Analytic       Model        (SAM)       {\sc       morgana}
(\citealt{2007MNRAS.375.1189M}).   In  SAMs,   the  evolution  of  the
baryonic component is followed by means of approximate, yet physically
grounded, analytic prescriptions for  modelling the relevant processes
(such  as  gas  cooling,  star   formation  and  feedback)  and  their
interplay, as a function of  the physical properties of model galaxies
(like   their  cold   gas   and  stellar   content).  These   analytic
prescriptions involve  a number  of parameters, usually  calibrated by
comparing model  predictions with a  well defined set  of low-redshift
observations.  Despite (and thanks  to) this simplified approach, SAMs
have turned into a flexible and powerful tool to explore a broad range
of  specific  physical assumptions,  over  scales  that could  not  be
directly modelled  simultaneously (ranging  from the accretion  onto a
super-massive black hole  on sub-pc scales to the  Mpc scales involved
in cosmological structure formation). In the following we will briefly
describe the treatment of the most relevant processes leading to bulge
formation in  {\sc morgana} (see \citealt{2011MNRAS.414.1439D},  for a
discussion of the different channels  for bulge formation in different
SAMs).

{\sc morgana}  distinguishes between  minor and major  galaxy mergers;
the threshold of the two events  being defined by a mass ratio between
secondary and primary galaxy of 0.3.  The orbital decay of dark matter
subhaloes and galaxy  mergers are modelled using  the fitting formulae
defined by \citet{2003MNRAS.341..434T}. In case of a minor merger, the
stellar mass and  the cold gas of the secondary  galaxy are completely
given to the bulge component of  the remnant galaxy, while the disc is
considered unaffected.  On the other  hand, in case of  major mergers,
the  whole  stellar and  gaseous  disc  of  the merging  galaxies  are
destroyed and  relaxed into a  spheroidal remnant. In both  cases, any
cold  gas eventually  associated  with the  bulge  can be  efficiently
converted  into   stars,  on   very  short   time-scales  (effectively
triggering  a ‘starburst’).  At  later times,  the remnant  spheroidal
galaxy can grow a new disc,  if cooling processes are effective in the
parent halo.  In particular for this work, we make use of the standard
realization of {\sc morgana} defined in \cite{2011MNRAS.414.1439D}.

\subsection{Galaxy models for N-body simulations}
\label{sec:models}

We  use   the  method  described  in   \citet{2005MNRAS.361..776S}  to
initialize our  galaxies. Each  object is  composed of  five different
components: (i)  a cold gaseous  disc with  mass \Mcg, (ii)  a stellar
disc (\Mdisc), (iii) a stellar bulge (\Mb), (iv) a gaseous halo (\Mhg)
and (v) a dark matter halo (\Mdm).

The  gaseous and  stellar  disc have  an  exponential surface  density
profiles  and  their scale  lengths  (\rg  and \rd  respectively)  are
related via  $\rg = \chi \rd$,  with $\chi=1.5$.  We adopt  a sech$^2$
profile with  a scale height $z_0$  for the vertical structure  of the
stellar disc and the vertical velocity  dispersion is set equal to the
radial  velocity dispersion.   The vertical  structure of  the gaseous
disc is computed  by requiring balance between  the galactic potential
and the  pressure given by the  Equation of State (EOS);  the EOS also
fixes the temperature of the gas, rather then the velocity dispersion.
Finally we  adopt an Hernquist profile  \cite{1990ApJ...356..359H} for
both the stellar bulge and the dark matter halo. The stellar bulge has
scale length  \rb, while the  dark matter halo  is defined by  a scale
length \rs,  a concentration parameter  $c=\rvir/\rs$ and a  halo spin
$\lambda$.

To model the hot gaseous hale we followed the same parameterization of
\citet{2011MNRAS.415.3750M}. Namely we use the observational motivated
$\beta$-profile                        (\citealt{1976A&A....49..137C},
\citealt{1984ApJ...276...38J}, \citealt{1998ApJ...503..569E}):
\begin{equation}
\rho_{\rm                hg}(r)                =                \rho_0
\left[1+\left(\frac{r}{\rc}\right)^2\right]^{-\frac{3}{2}\beta}\;,
\end{equation}
We  use  $\beta   =  2/3$  \citep{1984ApJ...276...38J},  $\rc=0.22\rs$
\citep{1998ApJ...497..555M}  and fix  $\rho_0$ such  that the  hot gas
mass within the virial radius is $M_{\rm hg}$.
The temperature profile is  fixed by imposing hydrostatic equilibrium.
In addition, we impose a slow  rotation around around the spin axis of
the discs for  the hot halo. The specific angular  momentum is defined
as $j_{\rm hg}$  and is set to  be a multiple of  the specific angular
momentum of the dark matter halo $j_{\rm dm}$ such that $ j_{\rm hg} =
\alpha j_{\rm dm}$.

High           resolution           cosmological           simulations
\citep[][e.g.]{2010Natur.463..203G}  have indicated  that this  ``spin
factor''  $\alpha$  is  generally   larger  than  unity,  as  feedback
processes preferentially remove low angular momentum material from the
halo.    For  the   exact  value   we   have  used   the  results   of
\citet{2011MNRAS.415.3750M},  who obtained  $\alpha=4$ using  isolated
simulations of  a MW-like galaxy  and requiring that the  evolution of
the  average stellar  mass and  scale-length found  observationally be
reproduced.

\subsection{Simulations}

The hydrodynamical  simulations have been performed  with the parallel
TreeSPH-code  {\sc  GADGET-2} \citep{2005MNRAS.364.1105S}.   The  code
uses Smoothed Particle  Hydrodynamics (\citealt{1977AJ.....82.1013L} ;
\citealt{1977MNRAS.181..375G}   ;  \citealt{1992ARA&A..30..543M})   to
evolve    the    gas    using    an    entropy    conserving    scheme
\citep{2002MNRAS.333..649S}. The code includes Radiative cooling for a
primordial    mixture    of     hydrogen    and    helium    following
\citet{1996ApJS..105...19K} and  a spatially  uniform time-independent
local UV background \citep{1996ApJ...461...20H}.
We model Star formation and  the associated heating by supernovae (SN)
following  the  sub-resolution  multiphase   ISM  model  described  in
\citet{2003MNRAS.339..289S}. {\bf }
Cold clouds form  stars in dense ($\rho>\rho_{th}$) regions  on a time
scale  chosen to  match observations  \citep{1998ApJ...498..541K}. The
threshold  density  $\rho_{th}$  is  determined  self-consistently  by
demanding that the equation of state  (EOS) is continuous at the onset
of star formation.

We    also     include    SN-driven    galactic     winds    following
\citet{2003MNRAS.339..289S}.  The wind mass-loss rate is assumed to be
proportional to  the star formation  rate (SFR) $\dot M_w=  \zeta \dot
M_*$,  where  the  mass-loading-factor  $\zeta$  quantifies  the  wind
efficiency. We assume a constant wind speed model with $v_w=500 \kms $
(energy-driven wind).  We refer to Table \ref{t:smtparameters} for the
values of  further parameters (assuming  a Kroupa IMF)  describing the
multiphase feedback model.  We do  not include feedback from accreting
black holes (AGN feedback) in our simulations.

\subsection{Simulations of Semi-Analytic Merger Trees}
\label{trees}

 We use  the method  devised by \citet{2014MNRAS.437.1027M}  to combine  ({\sc   pinocchio})+SAMS   results  with   high   resolution
hydrodynamical simulations.  As  a first step we select  a merger tree
from the  {\sc pinocchio}  simulation, we then  use the  {\sc morgana}
semi-analytical model to predict the properties of the galaxies hosted
in each  branch of the tree. The prediction for the properties of the galaxies directly taken from the {\sc MORGANA } SAM include the  masses of the gas disc, stellar disc, stellar bulge, hot gaseous halo and the dark matter halo and the scale lengths of the  stellar disc and the stellar bulge.  These predictions are used  to create a
particle based realization  of the galaxy (made of  dark matter, stars
and hot  and cold gas,  see section \ref{sec:models}) at  the designed
started time of the simulation (here $z_i=1$). The number of particles in each component is decided by fixing $N_{\star} = 500,000$, which is the total number of stellar  particles in the final merger remnant. This then sets the mass of the star particles. We also impose that the mass of DM particle be $15$ times more massive than the stars and the gas particles to be 
$2$ times more massive. The typical mass of DM, gas and stars in our simulations are $1.0 \times 10^6 \ \rm{M_\odot}$, $1.4 \times 10^5 \ \rm{M_\odot}$ and $6.6 \times 10^4 \ \rm{M_\odot}$ respectively.  The softening length ($\epsilon_i$) of each component `i' of the galaxy is given by $\epsilon_i = 32 \ \rm{kpc} \ \times \sqrt{m_{part,i}/10^{10} \ \rm{M_\odot}}$, where $m_{part,i}$ is the mass the particles of the `i'th  component. Typically for our simulations this equates to a softening of about about 100 pc for the gas particles, 300 pc for the DM particles and 80 pc for the stars.  

This galaxy  is then  evolved with the  hydrodynamical code  until the
time of  its first merger, as  predicted by the merger  tree.  At this
point we stop the hydro  simulations and create a particle realization
of  the satellite  system, which  is also  based on  the semi-analytic
prediction. We then add the satellite  in the simulation at the virial
radius of  the main halo and  we restart the hydrodynamical  run.  The
orbital parameters  (position and velocity  at the time  of accretion)
are  directly  taken  from  the   $N$-body  simulation,  in  this  way
``naturally'' creating  a cosmologically motivated merger.  The system
galaxy+satellite is  evolved with  the hydrodynamical code,  until the
next satellite galaxy  enters the main halo. This  process is repeated
for all  merging satellites until $z=0$.   Table \ref{t:smtparameters}
lists all parameters used to construct  and run our galaxies (see also
\citet{2014MNRAS.437.1027M} for further details  and the exact meaning
of all parameters).

\begin{table*}
 \centering
 \begin{minipage}{140mm}
  \caption
  {Summary of the parameters used  for the simulations of merger trees
    and their fiducial value.}  \null
  \begin{tabular}{@{}llr@{}}
  \hline Parameter & Description & Fiducial value\\ \hline
$z_i$ &  Redshift at  the start  of the  simulation &  1.0\\ $\mu_{\rm
    min}$ & Minimum  dark matter mass ratio & 0.03\\  $\delta$ & Ratio
  of scaleheight and scalelength of the stellar disc & 0.15\\ $\chi$ &
  Ratio of scalelengths between gaseous and stellar disc & 1.5\\ $\xi$
  & Ratio of gaseous halo core radius and dark matter halo scale radius
  &  0.22\\ $\beta_{\rm  hg}$  &  Slope parameter  of  gaseous halo  &
  0.67\\ $\alpha$ & Ratio of specific angular momentum between gaseous
  and  dark halo&  4.0\\  $N_*$  & Expected  final  number of  stellar
  particles in the  central galaxy & $200\,000$\\ $\kappa$  & Ratio of
  dark matter and  stellar particle mass & 15.0\\  $N_{\rm res,sat}$ &
  Ratio of satellite and central  galaxy particle mass & 1.0\\ $N_{\rm
    min}$  &   Minimum  number  of   particles  in  one   component  &
  100\\ $\epsilon_1$  & Softening length  in kpc for particle  of mass
  $m=10^{10}\Msun$ &  32.0\\ $t_0^*$  & Gas consumption  time-scale in
  Gyr  for  star  formation  model &  3.5$^\dagger$\\  $A_0$  &  Cloud
  evaporation  parameter for  star formation  model &  1250.0$^\dagger
  $\\  $\beta_{\rm SF}$  & Mass  fraction  of massive  stars for  star
  formation  model  &  0.16$^\dagger  $\\  $T_{\rm  SN}$  &  Effective
  supernova     temperature    in     K     for    feedback     model&
  $1.25\times10^{8\dagger}$\\ $\zeta$  & Mass loading factor  for wind
  model &  1.0\\ $v_{\rm wind}$  & Initial  wind velocity in  \kms for
  wind model  & 500.0\\ \hline \null\\  \multicolumn{3}{l}{$^\dagger $
    The star formation parameters assume a Kroupa IMF.}\\
\label{t:smtparameters}
\end{tabular}
\end{minipage}
\end{table*}

\section{Morphological evolution of galaxies}
\label{sec:sims}

As mentioned  earlier the most  important processes which  dictate the
morphology  of  galaxies  are  mergers  and  close  encounters.  These
processes are  known to trigger  mass transfer into the  central bulge
through different  channels; the  main ones being  : (i)  accretion of
satellite  material onto  the  bulge of  the  central galaxy  (central
bulge) during  a merger, (ii) the  transfer of stars from  the disc of
the  central  galaxy  (central  disc)   to  central  bulge  and  (iii)
funnelling of gas towards the centre and subsequent star formation.

As described  in Sec. \ref{subsec:sam}, SAMs  use simple prescriptions
for  mass transfer  through  these channels.  Are these  prescriptions
correct?  For example the simple assumption that during a major merger
all the  material of  the central  disc goes into  the bulge  has been
contradicted by \citet{2009ApJ...691.1168H}, who showed that discs can
survive major mergers if they are gas rich. \cite{2013MNRAS.tmp.1098C}
showed that a  disc dominated satellite is easily  tidally stripped as
it orbits a  halo before it finally mergers, which  reduces the amount
of material given by the satellite to the central bulge.

Here we test different  scenarios using high resolution hydrodynamical
simulations of  galaxy mergers,  starting from the  initial conditions
given by the semi-analytic model as described in the previous section.
We  simulate  a total  of  $20$  merger  events  covering a  range  of
different merger histories and galaxy properties. The merger trees and
their parameters  are given  in Table \ref{Table_trees}.   Finally for
each tree  we also perform  an 'isolation' run,  i.e. a run  where the
central  galaxy is  evolved  without  any mergers  from  $z=1$ to  the
present time.  The isolated runs act as control, which helps to easily
dis-entangle the effect of mergers.

\begin{table*}
 \centering
\begin{minipage}{120mm}
      \null
      \caption
  {Table listing the properties of simulated merger trees }
  \begin{tabular}{lcccccccccccccccccccc}
  \hline Tree/Sat  ID & $z_{enter}$\footnote{Redshift of  entry of the
    halo}    &   $\mu$\footnote{Dark    Matter    merger   ratio}    &
  $\mu_b$\footnote{Baryon  merger  ratio}  &  $log(M_h)$\footnote{Dark
    Matter    mass}    &     $log(M_*)$\footnote{Stellar    mass}    &
  $log(M_{cg})$\footnote{Cold gas  mass} & $\eta$\footnote{Circularity
    parameter} \\ \hline \hline

Tree 18989 & 1.0 &  - & - & 11.70 & 9.92 & 10.06 &  -\\ Sat 1 & 0.98 &
0.37 & 0.83 &11.27 & 9.92 & 9.65 & 0.20 \\ \hline Tree 28678 & 1.0 & -
& - & 11.67 &  10.44 & 9.62 & -\\ Sat 1 & 0.46 &  0.56 & 0.16 &11.47 &
9.66 & 8.74 & 0.44 \\ \hline Tree 80891  & 1.0 & - & - & 11.86 & 10.61
& 9.65 & -\\ Sat  1 & 0.81 & 0.05 & 0.037 &10.67 &  8.80 & 8.92 & 0.81
\\ \hline Tree 65521 & 1.0 & - & - & 11.81 & 10.48 & 10.21 & -\\ Sat 1
& 0.77 & 0.10 & 0.05 &10.84 & 9.05 & 8.93 & 0.43 \\ \hline Tree 154448
& 1.0 & - & - & 11.631 & 10.60 & 9.83 & -\\ Sat 1 & 0.67 & 0.97 & 0.74
& 11.651 & 10.74 &  8.79 & 0.42 \\ \hline Tree 215240 & 1.0  & - & - &
11.54 & 9.84 & 9.88 & -\\ Sat 1 &  0.37 & 0.76 & 0.47 & 11.52 & 9.86 &
8.21 & 0.13 \\ \hline Tree 455141 & 1.0 & - & - & 11.39 & 10.14 & 9.54
&  -\\ Sat  1 &  0.56 &  0.88 &  0.38 &  11.45 &  9.48 &  9.54 &  0.56
\\ \hline Tree 114590 & 1.0 & - & -  & 11.15 & 9.16 & 8.96 & -\\ Sat 1
& 0.92 & 0.86 & 0.84 & 11.10 & 9.12 & 9.32 & 0.71 \\ \hline Tree 28837
& 1.0 & - & - & 11.86 & 10.42 &  9.94 & -\\ Sat 1 & 0.72 & 0.04 & 0.01
& 10.52  & 8.48 & 7.33  & 0.31 \\  \hline Tree 50967 &  1.0 & - &  - &
11.86 & 10.34  & 10.17 & -  \\ Sat 1 & 0.45  & 0.062 & 0.05  & 10.69 &
8.46 & 8.52 & 0.43 \\ \hline Tree 52201  & 1.0 & - & - & 11.64 & 10.01
& 10.03 &  - \\ Sat 1 &  0.34 & 0.66 &  0.51 & 11.53 & 9.76  & 8.825 &
0.75 \\ \hline  Tree 53334 & 1.0  & - & - &  11.78 & 10.23 &  9.58 & -
\\ Sat 1 & 0.72  & 0.45 & 1.0 & 11.47 & 10.22 &  9.38 & 0.82 \\ \hline
Tree 58811 & 1.0 &  - & - & 11.78 & 10.36 & 9.65 &  -\\ Sat 1 & 0.17 &
0.38 & 0.3 & 11.40 & 9.55 & 9.49  & 0.15\\ \hline Tree 60367 & 1.0 & -
& - & 11.83 & 10.47 & 9.98 & -  \\ Sat 1 & 0.39 & 0.03 & 0.007 & 10.41
& 8.40  & 7.42 & 0.77  \\ \hline Tree  61557 & 1.0 &  - & - &  11.78 &
10.40 & 10.00 & - \\ Sat 1 & 0.28  & 0.34 & 0.13 & 11.37 & 9.53 & 8.86
& 0.14 \\ \hline Tree 102663 & 1.0 & - & - & 11.92 & 10.76 & 10.10 & -
\\ Sat 1 & 0.79  & 0.27 & 0.09 & 11.37 & 9.06 &  9.62 & 0.51 \\ \hline
Tree 350 &  1.0 & - & -  & 11.90 & 10.77 &  9.84 & -\\ Sat 1  & 0.96 &
0.13 & 0.041  & 11.02 & 9.25  & 8.67 & 0.35 \\  Sat 2 & 0.95  & 0.04 &
0.016 & 10.53  & 8.75 & 8.52 & 0.21  \\ Sat 3 & 0.93 &  0.04 & 0.009 &
10.62 & 8.67 & 7.73 & 0.36 \\ \hline Tree 2536 & 1.0 & - & - & 11.63 &
10.36 & 10.15 & -\\ Sat 1 & 0.79  & 0.12 & 0.025 & 10.75 & 8.76 & 8.28
& 0.69 \\  Sat 2 & 0.75  & 0.29 & 0.312 &  11.20 & 9.98 &  6.20 & 0.66
\\ Sat 3 & 0.46 & 0.19 & 0.135  & 11.18 & 9.46 & 8.94 & 0.31 \\ \hline
Tree 187460 & 1.0 &  - & - & 10.94 & 9.14 & 9.21 &  -\\ Sat 1 & 0.70 &
0.80 & 0.85 & 11.09 & 9.17 & 8.82 & 1.05 \\ Sat 2 & 0.61 & 0.66 & 0.79
& 11.32 &  9.26 & 8.10 &  0.63 \\ \hline Tree 159419  & 1.0 & -  & - &
11.05 & 9.08 & 8.26 & -\\ Sat 1 &  0.89 & 0.52 & 0.41 & 10.79 & 8.75 &
7.80 & 0.18 \\ Sat 2 & 0.88 & 0.77 & 0.66 & 11.15 & 9.34 & 8.15 & 0.49
\\ \hline \hline


\end{tabular}
\label{Table_trees}

\end{minipage}
\end{table*}

\subsection{Bulge/Disc Decomposition}
\label{sec:btd}
There are many  different ways to decompose the mass  of a galaxy into
its basic morphological entities i.e.,  disc and bulge, for example by
fitting surface  brightness profiles with  a bulge and  disc component
(e.g., \citealt{2006ApJ...645..986R}),  kinematic decompositions based
on  one  or  two  dimensional kinematic  maps  and  three  dimensional
component  fits.  All  these  methods  rely  heavily  on  the  viewing
angle. In this paper we make use of the six dimensional phase space of
($\rm{x, y, z, v_{x}, v_{y}, v_{z}}$), to track the bulge and the disc
components   of  the   primary  galaxy,   throughout  the   simulation
(see also  \citealt{Sc2010, Marinacci2014}). The  galaxy in
question is  viewed edge on i.e.,  the angular momentum vector  of the
disc is  placed parallel to  the $z$ axis.  Now for a  purely rotating
disc the circular  velocity of a given particle, at  a distance $r$ is
given by

\begin{equation}
v_{c} = \sqrt{\frac{GM(<r)}{r}}
\label{circ}  
\end{equation}  
and the specific angular momentum of the particle will be
\begin{equation}
l_{c} = r\sqrt{\frac{GM(<r)}{r}}
\label{circl}  
\end{equation}

If a particle  is purely rotationally supported then the  ratio of its
specific angular  momentum in  the $z$ direction ($l_z$)  and $l_{c}$ ($f_{rot}$)  will be
equal to $1$.
 \begin{equation}
f_{rot} = \frac{l_{z}}{l_c}
\label{lrot}  
\end{equation}

\begin{figure*}
 \subfigure[]{
   \includegraphics[scale=0.60]{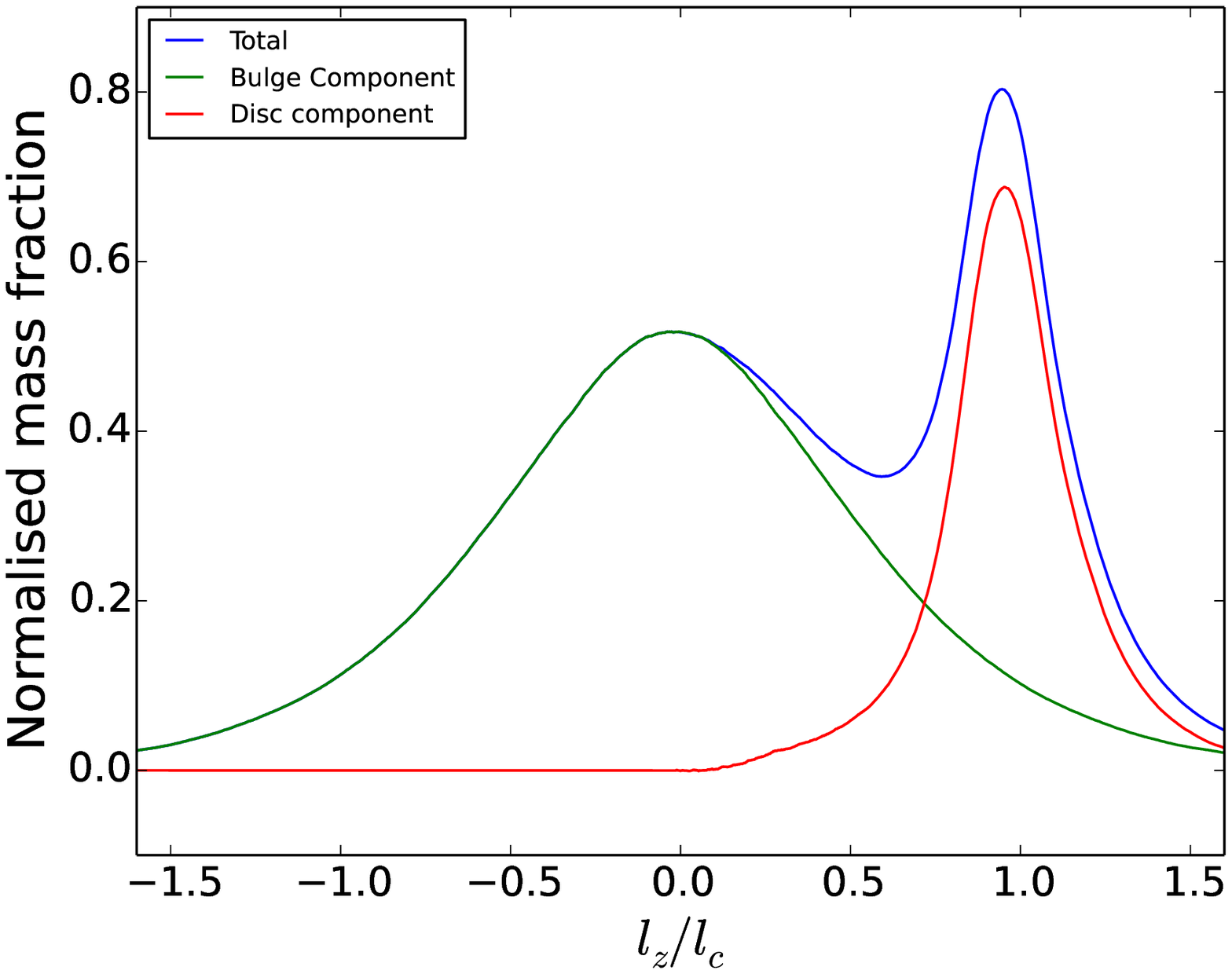}
   \label{fig:bdd}
   } 
   \subfigure[]{
   \includegraphics[scale=0.51]{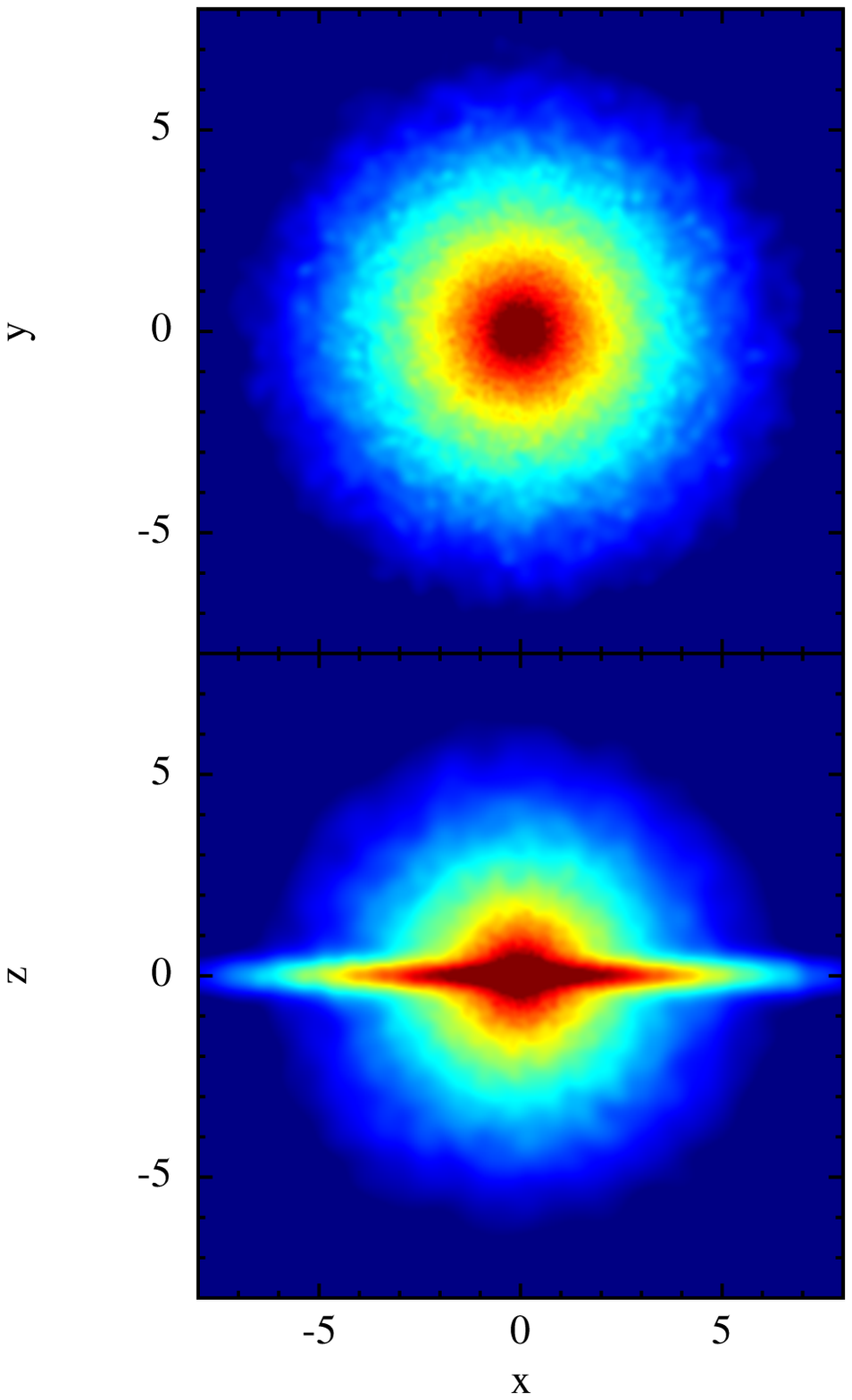}
   \label{fig:xyzview}
 }

 \caption[Optional   caption   for    list   of   figures]{An   sample
   decomposition of a bulge dominated galaxy in our simulations. \\ a)  The mass distribution  of the stellar component
   of the  galaxy as a  function of  the rotational support.  There is
   clearly a bi-modal distribution,  with clear difference between the
   bulge   (green)  and   disc  (red)   components.  This   particular
   decomposition     yields     a     $B/T     \approx     0.7$. \\ b)The column  density of the stellar  component shown in
   projection,  perpendicular  and  parallel   to  the  total  angular
   momentum           axis          of           the          galaxy. }
\end{figure*}

\begin{figure*}
 \centering \subfigure{ 
 \includegraphics[scale=0.42]{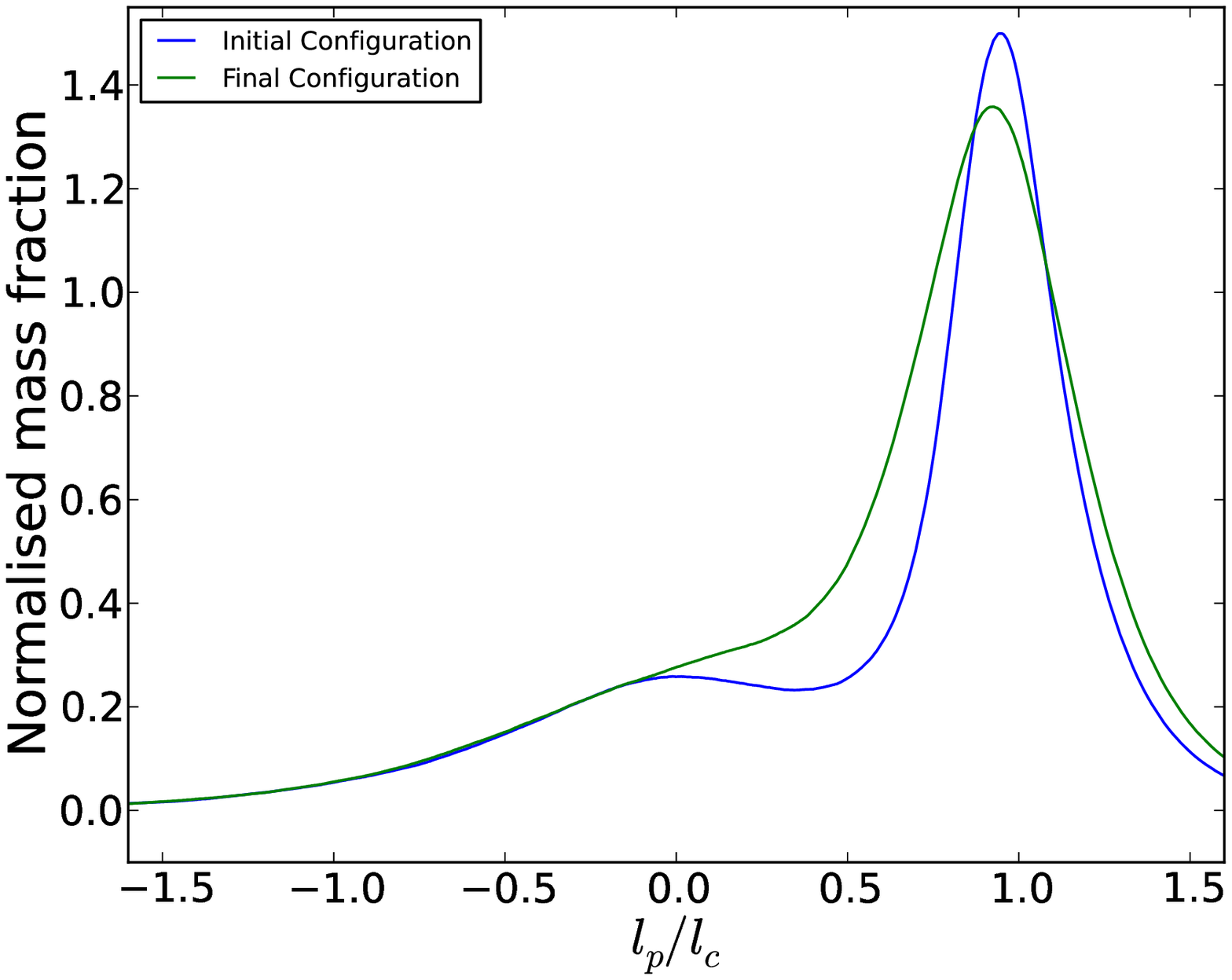}
   \label{fig:min}
   } 
   \subfigure{
    \includegraphics[scale=0.42]{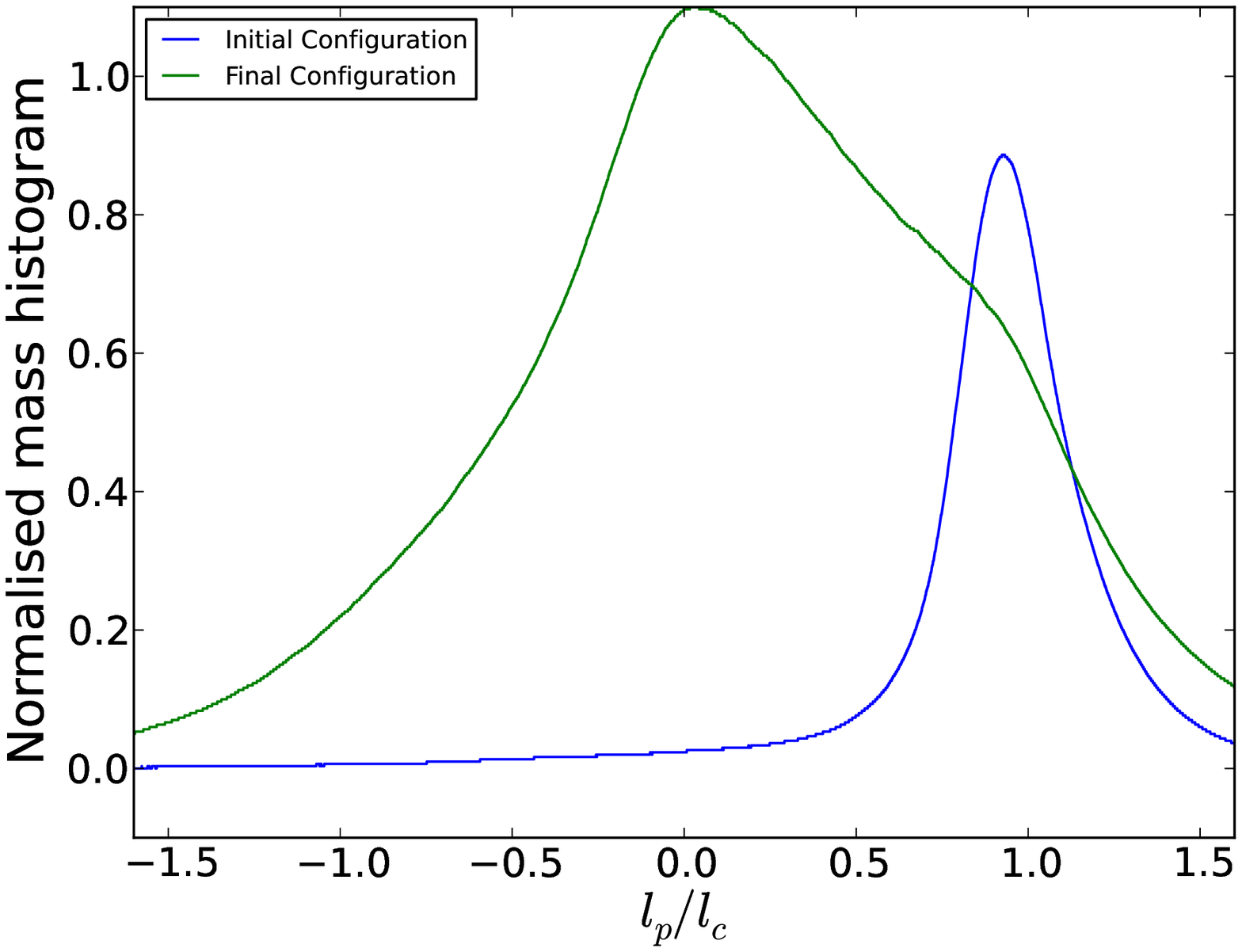}
   \label{fig:maj}
 \label{fig:minmaj}
 }
 \caption[Optional caption  for list of figures]{The  angular momentum
   distribution  of   the  central   galaxy  before(blue   curve)  and
   after(green curve)  a merger event.  Left Panel: Change  in stellar
   mass distribution  of the  central galaxy  after undergoing  a 1:10
   merger. The merger heats up the thin disc, leaving the bulge almost
   completely  unaltered.   Right  Panel:   Change  in   stellar  mass
   distribution  of  the central  galaxy  after  undergoing two  major
   mergers (1:1.25  $\&$ 1:1.5).  Due to the  high merger  ratios, the
   galaxy builds up  a large bulge, with  only a very small  hint of a
   remnant disc. }
\end{figure*}

 If we consider all the stellar component of a sample galaxy
from  our simulations  and  plot the  distribution of  the  mass as  a
function    of     the    rotational    support     i.e.,    $l_z/l_c$
(Fig.  \ref{fig:bdd}),   we  expect  it   to  be  bi-modal   with  the
rotationally  supported  disc  particles distributed  around  $f_{rot}
\approx  1$ and  the  velocity dispersion  ($\sigma$) supported  bulge
particles grouping around $f_{rot} \approx 0$. Decomposing the distribution into bulge and disc components comprises of figuring out the local maxima  close $f_{rot} = 0$. In the simple example shown in Fig.  \ref{fig:bdd} $f_{rot}$ is symmetrical
  around $0$, meaning  that the bulge is non-rotating.  This might not
  be true for many galaxies where the bulge might have some low amount
  of rotation  induced during a  merger (for  e.g. see right  panel of
  Fig. \ref{fig:minmaj}).  In order to make a  self consistent
  decomposition,  instead  of  assuming the  symmetric nature of  $f_{rot}$ around $0$, we  calculate the point at which the
  slope or first derivative of  the stellar distribution function
  becomes zero, particularly points where the slope changes from positive to negative (i.e. local maxima). This takes  care of cases where the bulge
  has  a rotation,  but this  only works  if the  bulge and  disc 
  distributions are  quite separated  from each other. In  some galaxies
  the two distributions  merge and the function no longer  has a local
  maxima.  This  problem is  overcome  by  calculating the  inflection
  points  of  this  function, especially the points where the second derivate changes sign from negative to positive (i.e. the curve changes from convex to concave). This neatly separates
  the bulge and disc components even in galaxies where the stellar and
  bulge distributions overlap quite a bit. It should be mentioned that
  we look  for these local  maxima and inflection points  only between
  $-0.3  \le f_{rot}  \le 0.3$,  beyond  which we consider that the galaxy has no bulge component.  The stars in the bulge are then assumed to be symmetrically distributed around the local maximum or inflection point, while the disc stars make up the rest of the distribution (as shown in  Fig. \ref{fig:bdd}).  This technique, by construction has the ability to only statistically determine the bulge and disc components of a simulated galaxy and cannot uniquely classify individual particles as belonging to either of the components. For example in Fig. \ref{fig:bdd} all particles with $f_{rot} \sim 0.7$ have equal probability of being a bulge and disc particle.

   This decomposition on a
sample galaxy  at redshift z=1,  from our  catalog (Tree 350)  gives a
$B/T$ $\sim$ $0.7$, which is confirmed by the initial conditions taken
from  the SAM.   The stellar  density maps  (figure \ref{fig:xyzview})
also confirm that this is a bulge dominated galaxy.

Fig. \ref{fig:minmaj}  shows the  effect of a  minor and  major merger
events  and how  the angular  momentum distribution  of matter  in the
central galaxy  changes in our  simulations.  A minor merger  does not
particularly affect  the morphology of the  galaxy (Fig.\ref{fig:min})
which  remains disc  dominated  (Tree  65521).  On  the  other hand  a
multiple  major merger  completely  changes the  galaxy morphology  as
shown in  of Fig.  \ref{fig:maj}. A disc  dominated galaxy  turns into
bulge dominated  one with a  very small hint  of a remnant  disc (Tree
187460).

\begin{figure*}
\includegraphics[scale = 0.7]{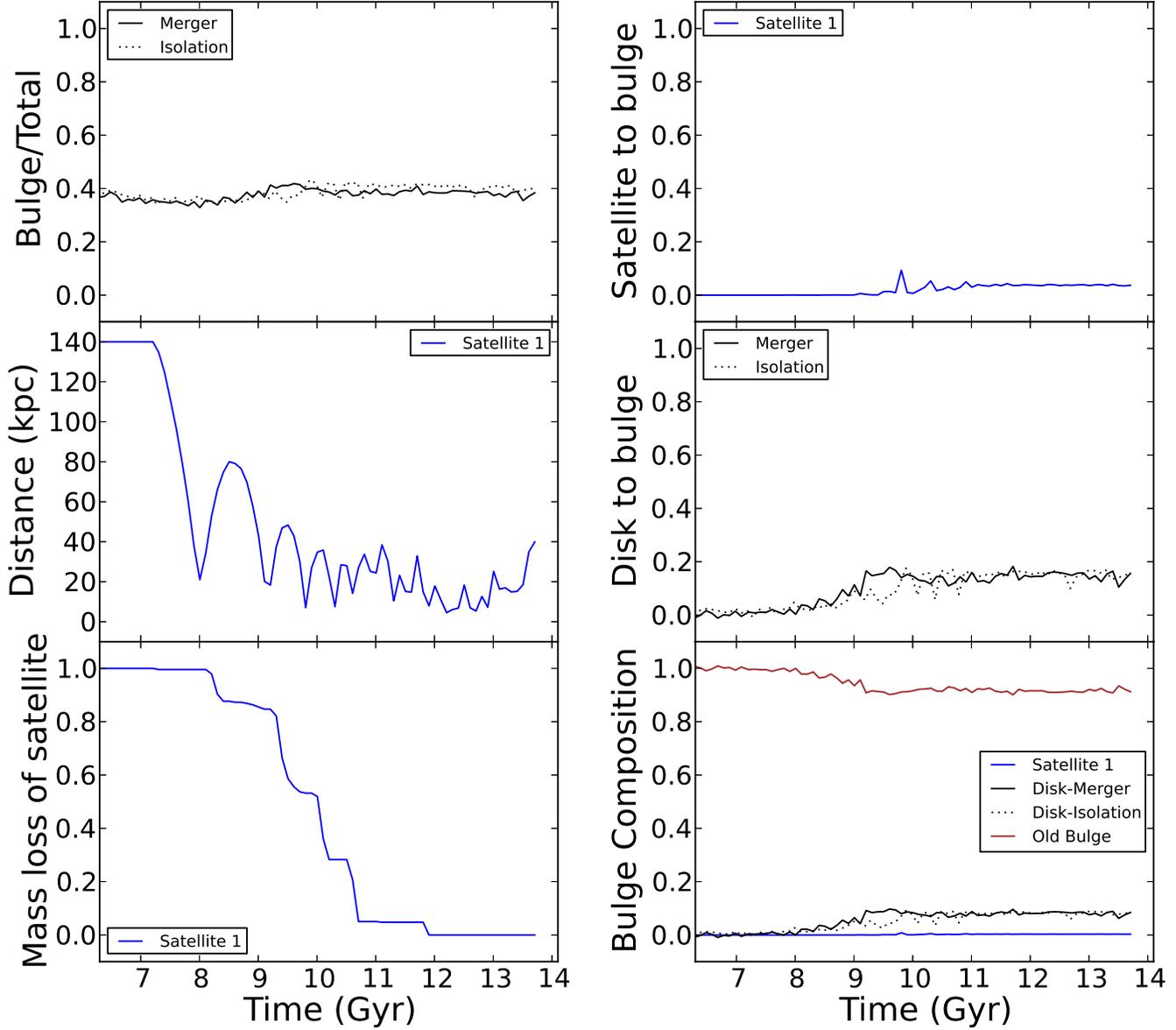}
\caption{Time evolution  of merger parameters  for Tree 65521  ($\mu =
  0.10$).  Panel 1 - Evolution of  $B/T$ in both the isolated (dotted)
  and merger(solid) runs, Panel 2 - Distance of the satellite form the
  centre of  the primary galaxy,  Panel 3 -  Mass loss of  the stellar
  component of the  satellite as it orbits in the  primary halo of the
  central galaxy,  Panel 4 -  Stellar mass  of the satellite  given to
  bulge of the  primary galaxy, Panel 5 - Amount  of initial disc mass
  given to the bulge, (during a  merger: solid curve and in isolation:
  dotted curve) and Panel 6 - Composition of the bulge. The brown line
  denotes the fraction of stars which  existed in the central bulge at
  the start of the simulation, the  blue line is the contribution from
  the  satellite  and the  black  (solid  and  dotted) lines  are  the
  contribution  from  the   central  disc  during  a   merger  and  in
  isolation. }
\label{fig:minall}
\end{figure*}

\begin{figure*}
\includegraphics[scale = 0.7]{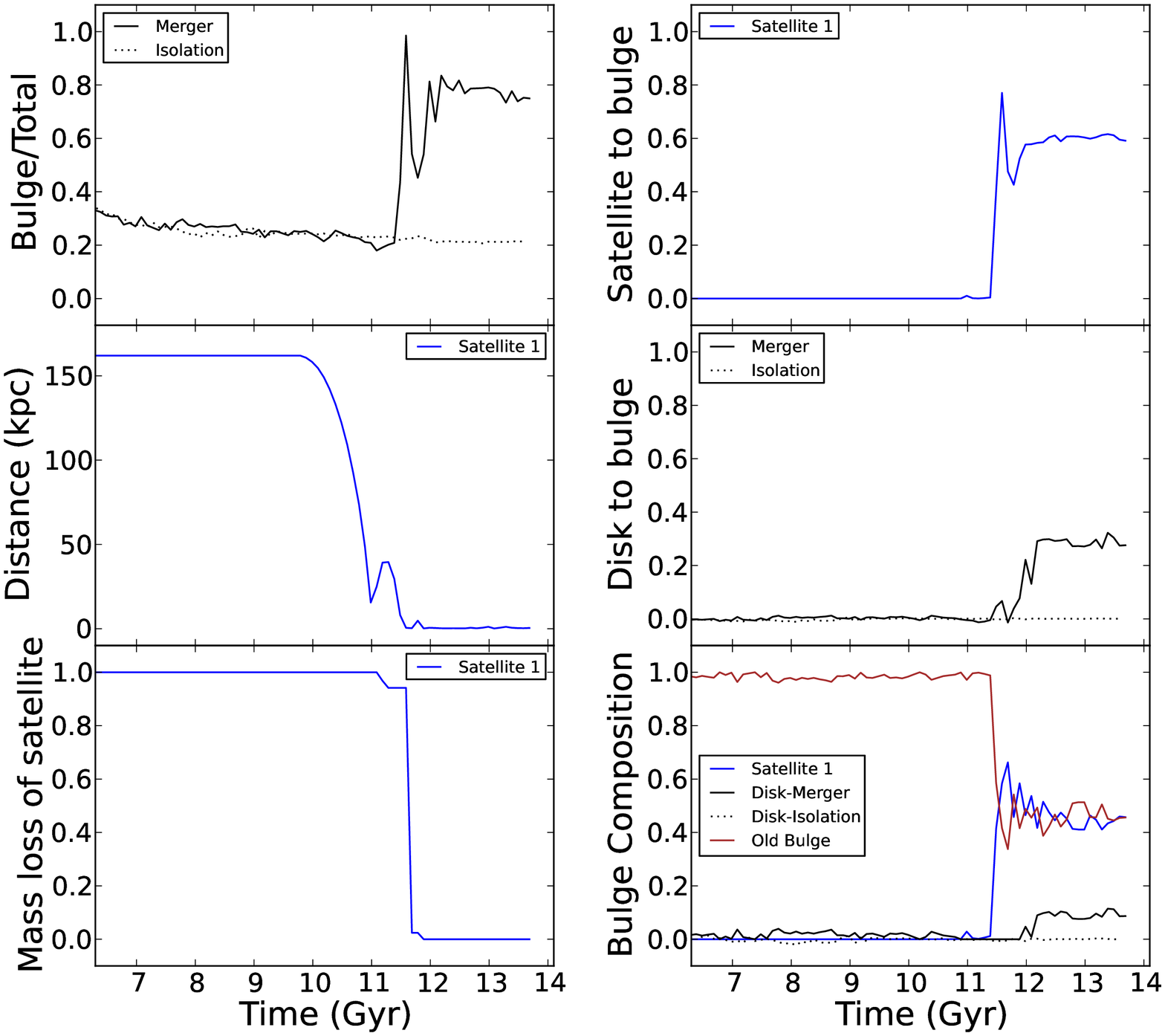}
\caption{Time evolution  of merger parameters  for Tree 18989  ($\mu =
  0.37$).  The  panels  denote  the   same  quantities  as  in  Figure
  \ref{fig:minall}}
\label{fig:majall}
\end{figure*}

This method  also allows us the  track the mass flow  into the central
bulge during the  whole simulation, as shown  in Fig. \ref{fig:minall}
(Tree 65521;  $\mu =  0.1$) and \ref{fig:majall}  (Tree 18989;  $\mu =
0.76$). Every particle in  a {\sc gadget-2} simulation has
  a unique  ID, and for  DM and  stars, this ID  does not change  as a
  function of simulation time. When we set up the initial conditions we decompose
  the galaxy into bulge/disc components and  make a note of the IDs of
  the particle in  each component. As the simulation  evolves in time,
  the particles are tracked through their unique IDs as they move from
  one component  to other. This  allows us  to track the  various mass
  transfer channels very efficiently. Since  we only look at particles
  that are present at the start of  our simulation we do not track how
  the morphology of the newly formed stars changes.
  
We track  the stars of  the satellite in  the same way  and we
  consider that the particle has  been unbounded from the satellite if
  the particle  is at a distance  $>3r_s$, or lies beyond $5 \kpc$  above  or below  the disc. Some  extreme
  galaxies might have a massive bulge and might have bulge stars above
  the $5 \kpc$ limit  that we have used here. In order  to see if this
  classification is robust we looked at satellite stripping with
  the stripping radius of $7.5 \&  10.0 \kpc$ above and below the disc
  of the galaxy. In all the simulations of our sample we get less than
  $10 \%$  changes in all related  merging channels when we  make this
  change. This is because even if  there is some amount of bulge stars
  present beyond this  height, the density of stars is  very low. This
  makes  the error  on  the  channels also  quite  low. Therefore,  we
  continue to use this definition throughout the paper.

Figures \ref{fig:minall} $\&$  \ref{fig:majall}  show  the evolution of various properties  as a function
of  simulation time.  The top  left panel  in these  figures show  the
evolution  of the  bulge/total ratio  of  the central  galaxy in  both
merger and isolation. During a minor merger the difference between the
B/T ratio  is minimal whereas when  a galaxy undergoes a  major merger
the B/T  ratio increases  considerably. The middle  panel on  the left
shows the distance of the satellite  to the central galaxy. Due to the
dependence of the merging time on the  mass ratio of the merger, a low
mass  satellite  spends   more  time  orbiting   the  primary
galaxy.  The bottom  panel on  the  left shows  the mass  loss of  the
satellite. The low mass satellite undergoes numerous close encounters,
losing  most of its  mass to  the central  halo in  the process,
whereas the high  mass satellite has fewer orbits  and deposits almost
all of  its mass into  the central bulge. The  top panel on  the right
shows the amount of satellite stellar  mass given to the central bulge
and it is  seen that a minor satellite deposits  a very small fraction
of its stellar mass into the central bulge owing to the fact that most
of its mass is  stripped away and lost to the  halo during its descent
while a  massive satellite deposits  almost all  of its mass  into the
central bulge as it  spends less time orbiting and it  is also able to
hold on to more of its mass  before it merges. The middle panel on the
right shows the  evolution of the amount of central  stellar disc mass
given to the central bulge during  a merger and in isolation. There is
almost no mass transfer from the  central disc to the central bulge in
case of a minor merger, but during  a major merger about $30\%$ of the
central stellar disc ends up in the central bulge. This number is very
small compared to most SAM  prescriptions which assume that almost all
the stellar  mass of the  central disc ends up  in the bulge  during a
major merger. As we will discuss later a large fraction of the central
disc gets  dispersed into the stellar  halo, and a small  remnant disc
survives. Finally, the bottom panel on the right shows the composition
of the central bulge.
As expected  the low mass  satellite contributes little, while  a high
mass satellite  contributes a  large amount of  material to  the bulge
and, in the end, it dominates the bulge composition in this particular
example.

In the following  sections we look closer to the  various channels for
mass transfer  into the bulge  and we will  also try to  quantify them
with simple, physically motivated, empirical relations.

\subsection{Importance of merger parameters}

The major  parameters which control the  outcome of a merger  event is
the ratio  of masses  of a  merging galaxies and  the their  ration of
orbital angular  momentum. The  ratio of  the total  halo mass  of the
galaxies  ($\mu$)  will control  the  dynamics  of infall,  while  the
components which  dominate the  potential at the  center of  the halos
(stars +  cold gas) will provide  most of the torque  during the final
stages of the merger. Hence, we  can expect the baryonic (stars + cold
gas) merger  ratio ($\mu_b$) to be  important as well. In  addition we
also the dependence of various transfer channels on $\mu_c$, the ratio
of the  amount of  total material within  two scale  radii (calculated
here  assuming  that  the  DM  halo follows  a  NFW  profile)  of  the
halos. This parameter is chosen so as to compare our results to H09.

In   addition,  previous   studies  \citep{2013MNRAS.tmp.1098C}   have
stressed  the  importance  of  the  satellite  orbital  parameters  in
determining  the final  outcome  of  the merger.  In  our approach  we
extract our merger trees and the satellite orbital parameters directly
from  cosmological   simulations.  This   ensures  that   the  orbital
properties  of  satellites  fall  within the  gamut  of  the  normally
occurring merging paths in the Universe. 

The drawback of
  this  method  is  that, there is no control  over  the  input
  parameters. This is makes it difficult to make a controlled experiment on the impact of various merger parameters (see for example the GALMER data base; \citealt{Galmer}).
  However it is still important to check the validity  of  previous
  works in  a  full  cosmological setting as we do.  However, we do try to quantify the effects of the mergers on the galaxy morphology as a function of various parameters.
  Our analysis is not meant to fully reproduce the dependence of mass transfer on the merger properties (our sample is
too small for such a task), but to grasp the main trends and dependencies and provide
guidance to test the implications of our findings in broader context.

We look at the most important factors that could affect the mergers. A
high mass satellite experiences a  high dynamical frictional force and
will fall quickly  to the center, thereby depositing most  of its mass
into the  bulge of the central  galaxy.  The opposite is  true for low
mass mergers.  The satellite  orbits around  the central  galaxy many
times, losing  its mass  through repeated  close encounters  before it
merges.  The circularity  of the orbit can be expressed  as a function
of the  ratio between  the orbital angular  momentum of  the satellite
($j$) and $j_{c}(E)$ the angular momentum of a circular orbit with the
same energy as the satellite, i.e.,
\begin{equation}
\eta = \frac{j}{j_{c}(E)}.
\end{equation}
Of  course the  mass transfer  in  various channels  will be  directly
proportional to  the merger ratios  $\mu$, $\mu_b$ and/or  $\mu_c$ and
inversely proportional to $\eta$. We  quantify the dependence on orbit
by  $\exp  (1.9  \eta)$,  where   the  exponential  function  and  the
proportionality factor of $1.9$ comes from the relation between $\eta$
and   the    total   merger    time   for   satellites    derived   by
\cite{2008MNRAS.383...93B} from cosmological  simulations. We test the
dependence  of   mass  transfer   through  various  channels   on  the
parameters.  We use a very simple functional form

\begin{multline}
f(x)  =  ax  \\  where  \  \ x  \  \epsilon  \  \{\mu,  \mu_b,  \mu_c,
\frac{\mu}{exp(1.9\eta)},                                       \frac{
  \mu_b}{exp(1.9\eta)},\frac{\mu_c}{exp(1.9\eta)}\}
\end{multline}

which ensures that  the boundary value condition of  $f(x)=0$ at $x=0$
is satisfied.   For each  of the  mass transfer  channels we  test its
dependence on  all the  four parameters and six functional forms. Table  \ref{Table_chi2} shows
the best fit value of $a$ for all the parameters and it also gives the
corresponding $\chi^2$.   We see that the results are not very different for the different
merger parameters; moreover, a more sophisticated analysis designed to
differentiate between the parameters is not feasible given our limited
sample. In fact, the $\chi^2$ analysis doesn't provide compelling evidence
in favour of adding parameters and/or physical dependencies, with
respect to the simple assumption of $1$-parameter linear dependence from
a given physical property. Therefore, we just consider dependencies
from the parameter providing the overall smaller $\chi^2$ value, which in
our case happens to be $\mu_b$ , and plot the functional dependence of
the transfer channels on $\mu_b$ from now on.

\subsection{Where does the satellite mass end up?}

One of the major channels to build up the bulge of a galaxy is through
the transfer of  material from the satellite directly  into the bulge.
The {\sc morgana} model employs a simple set of formulae to evolve the
bulge mass  during mergers. For example  it is assumed that  the whole
satellite mass is  added to the bulge, irrespective of  the mass ratio
between the satellite and the host and/or the orbit of the satellite.

\begin{figure}
 \includegraphics[scale = 0.45]{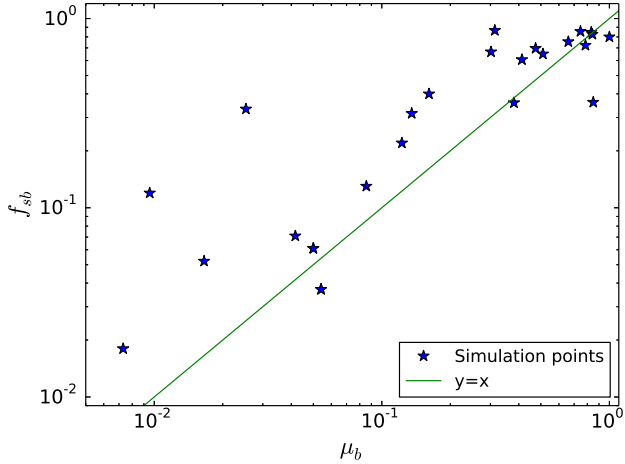}
\caption{The fraction of satellite mass  given to the central bulge as
  a function of  their baryonic merger ratios. The green  curve is out
  fit to the simulated data (blue stars).}
\label{fig:fsb}
\end{figure}

We  define $f_{sb}$  as the  fraction of  the initial  (before infall)
stellar  mass of  the satellite  given  to the  central bulge.  Figure
\ref{fig:fsb}  show  the  results  for  $f_{sb}$  from  our  suite  of
simulations (blue  stars) as a  function of $\mu_b$.   The simulations
data show a significant dependence  of $f_{sb}$ on the baryonic merger
ratio.

The best fit parameters that  describe $f_{sb}$ in our simulations are
given by:
\begin{equation}
f_{sb} = 1.0\mu_b
\label{fitfsb}
\end{equation} 
represented by the green line in figure \ref{fig:fsb}, which is a fair
representation of  the simulations results (also  matches results from
H09), while being still simple to implement in semi-analytic models.

\subsection{Where does the central disc mass end up?}

Another major  channel to build up  the bulge is the  transfer of mass
(stars+gas) from  the central disc  into the central bulge  during and
after a  merger.  In the {\sc  morgana} code the following  formula is
used  to  determine  the  mass  ($M_{db})$  transferred  through  this
channel:
\begin{equation}
\frac{M_{db}}{M_{disc}} = \left\{
  \begin{array}{l l}
    0 & \quad \mu \le 0.3\\ 1 & \quad \mu > 0.3\\
  \end{array} \right.
\end{equation}
where $M_{disc}$  is the central  disc mass  (stars and gas).   On the
other  hand the  gaseous and  stellar  components of  the disc  behave
differently  as pointed  out  by H09.   They have  shown  that if  the
central disc is gas  rich then only a small amount  of its initial gas
disc is funnelled into the bulge.   This mass transfer is given by the
following relation
\begin{equation}
f_{cgb}= (1-f_{gas})\mu_c
\end{equation}
where  $f_{cgb}$  is  the  fraction  of  the  central  disc  gas  mass
transferred to the central bulge, and $f_{gas}$ is the gas fraction of
the   central   disc  (ratio   between   gaseous   and  stellar   disc
mass). $\mu_c$ is the central (within two scale radii) merger ratio as
defined by  H09 and  used in  many SAM  calculations (see  for example
\citealt{2012MNRAS.423.1992S} and \citealt{2014MNRAS.444..942P}).

\begin{figure}
\centering \includegraphics[scale = 0.45]{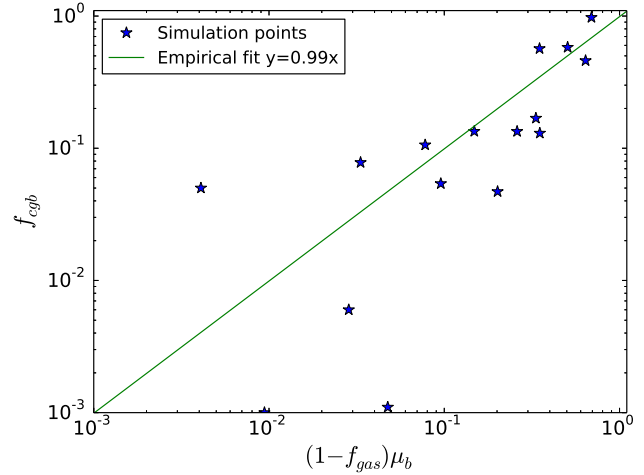}
\caption{The fraction  of central gas  disc mass given to  the central
  bulge. The blue points are simulation  results and the green line is
  the empirical fit to the points. }
\label{fig:fgb}
\end{figure}

In this  section we show  the results of  single merger events  in our
simulations inventory. The  effect of multiple mergers  on the central
disc is very hard to quantify. The mass transfer from the disc to bulge, seems to be dependent not only on the properties of the merging galaxies but also on the time lag between the mergers because it is easier to perturb an already unstable disc. Therefore, quantifying  the effect of multiple mergers is quite involved and we defer this to a future paper. In this work we only concentrate on the results from the  binary merger simulations, the results of which are shown in
Fig. \ref{fig:fgb}.  We find an  empirical relation for the  amount of
gas mass transfer into the bulge given by
\begin{equation}
f_{cgb}= (1-f_{gas})\mu_b
\label{eq:cgas}
\end{equation}
shown as  solid (green)  line in  figure \ref{fig:fgb}.  Therefore, we
confirm  previous results  on the  importance of  the gas  fraction in
determining the gas transfer from central disc to central bulge.

On the other hand we find quite a substantial difference with previous
studies for what concerns the fate of the central stellar disc after a
merger. Many SAMs assume that all the stellar mass of the central disc
ends up  in the central  bulge after a  major merger, while  H09 found
that this mass transfer is given by
\begin{equation}
f_{db} = \mu_c
\label{eq:fdbh} 
\end{equation}

\begin{figure}
\centering \includegraphics[scale = 0.45]{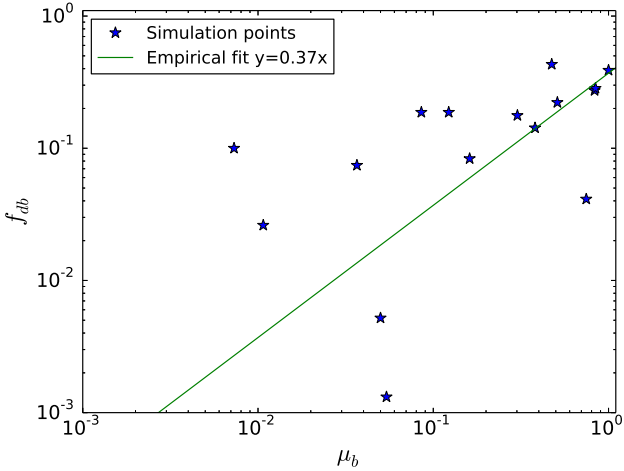}
\caption{The fraction of the central stellar disc mass that ends up in
  the  bulge of  the  galaxy  as a  function  of  the baryonic  merger
  ratio. The blue points are our simulation results and the green line
  is the fit to our result. }
\label{fig:fdb}
\end{figure}

However  in our  simulations even  for a  $1:1$ merger  the amount  of
central stellar  disc mass that ends  up in the central  bulge is only
about $37\%$,  as shown in  Fig. \ref{fig:fdb}.  The empirical  fit to
simulation data points is gives us a relation
\begin{equation}
f_{db} = 0.37\mu_b
\label{eq:fdb} 
\end{equation}
where $f_{db}$ is the fraction of  central stellar disc mass that ends
up into the bulge.

\begin{figure}
\centering \includegraphics[scale = 0.45]{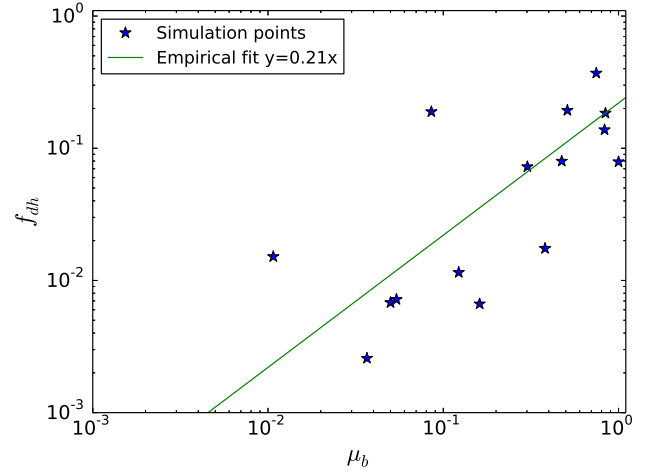}
\caption{The fraction of the central disc mass transferred to the halo
  as a function  of the baryonic merger ratio. The  green curve is out
  fit to the data.  A significant fraction of the stars  end up in the
  stellar halo of the central galaxy.}
\label{fig:fdh}
\end{figure}

This result is very interesting and diverges a lot with the simplest picture of major mergers
destroying discs, included in most (but not all) SAMs. Now the question arises:  where does
the rest  of the disc end  up? Major mergers are  very violent events,
where a lot  of energy (mainly orbital) is quickly  transferred to the
disc  stars.  As  a  consequence  a substantial  fraction  of disc  is
ejected into the stellar halo. In  our simulations the stellar halo is
defined as all  the stellar content present outside  three scale radii
of  the central  disc  and/or  $5 \kpc$  above  or  below it.   Figure
\ref{fig:fdh} shows the fraction of  central disc stellar mass that is
dispersed into  the stellar halo  of the galaxy  as a function  of the
baryonic merger ratio.

A disc destroyed during a major merger does not entirely end up in the
newly formed bulge but a significant fraction of its mass, up to $\sim
22\%$,  is ejected  into the  halo.  If  this mass  transfer from  the
central  disc to  the halo  is neglected  then the  bulge fraction  of
galaxies  will be  significantly  overestimated.  Recent results  from
\citet{2013MNRAS.436..697B}  have shown  that there  is a  substantial
amount of starlight in the extended envelopes of massive galaxies. Our
results show that  mergers might be one of the  ways to create massive
amounts of intercluster  light in these galaxies.  Empirically we find
that the fraction of central disc mass dispersed into the stellar halo
during a merger ($f_{dh}$)is given by
\begin{equation}
f_{dh} = 0.22\mu_b
\label{eq:fdh} 
\end{equation}

These results  show a  revised picture  of the the  fate of  a central
stellar disc during a major merger.  For a binary merger, about $40\%$
of the disc  mass loses angular momentum and is  then transferred into
the bulge.  More  than $20\%$ the initial mass gains  enough energy to
escape  from the  central region  and is  dispersed into  the galactic
halo. Finally a small  part of the disc is able  to survive the merger
and form  a smaller (thicker)  disc structure around the  newly formed
bulge.   These results  show  that  mergers are  not  as effective  as
previously thought in creating galactic bulges.

\begin{table*}
 \centering
\begin{minipage}{150mm}
      \null
      \caption
  {Table listing the best fit  parameters of the functional form $f(x)
    = \rm{a}x$ }
  \begin{tabular}{ | c | c | c | c | c | c | c | c | c | c | c | c | c |}
  \hline   Transfer  Channel   &   \multicolumn{2}{   c|  }{$\mu$}   &
  \multicolumn{2}{ c|  }{$\mu_b$} &  \multicolumn{2}{ c|  }{$\mu_c$} &
  \multicolumn{2}{ c| }{$\frac{\mu}{exp(1.9\eta)}$} & \multicolumn{2}{
    c  |   }{$\frac{\mu_b}{exp(1.9\eta)}$}  &  \multicolumn{2}{   c  |
  }{$\frac{\mu_c}{exp(1.9\eta)}$} \\ \hline

   - & a & $\chi^2$ & a & $\chi^2$ & a & $\chi^2$ & a & $\chi^2$ & a &
   $\chi^2$ & a  & $\chi^2$\\ \hline $f_{sb}$  & 0.92 & 1.36  & 1.01 &
   1.06  &  0.91 &  1.58  &  1.71  & 1.95  &  1.89  &  1.69 &  1.86  &
   1.77\\ \hline $f_{db}$ & 0.31 & 0.26 &  0.37 & 0.19 & 0.30 & 0.32 &
   0.54 & 0.25 & 0.58 & 0.26 & 0.56 & 0.29 \\ \hline $f_{dh}$ & 0.20 &
   0.10 & 0.22 & 0.11 & 0.20 & 0.12 & 0.31 & 0.14 & 0.35 & 0.13 & 0.34
   &  0.14  \\  \hline  $f_{cgb}$\footnote{All  the  functional  forms
     include a multiplicative factor of $(1-f_{gas})$} & 0.67 & 0.77 &
   0.99 & 0.29 & 0.64 & 0.86 & 1.23 & 0.86 & 1.82 & 0.43 & 1.29 & 0.85
   \\ \hline

\end{tabular}

\label{Table_chi2}
\end{minipage}
\end{table*}

\subsection{Sites of star formation}

There has been a lot of  effort put into quantifying the efficiency of
merger driven  starbursts (\citealt{2008MNRAS.384..386C},  H09, \citealt{Karman2015}).  H09 assume that  all the disc gas mass that enters
the bulge  during a merger  will be  available for star  formation and
there is a  particular efficiency for the conversion of  this gas into
stars.    \citet{2008MNRAS.384..386C}  have   looked  at   star  burst
efficiency  of the  entire  galaxy.  The SAMs  use  these results  for
starburst efficiency but they assume  that star formation mainly takes
place in the bulge of the galaxy.

Recent simulations  have shown  that there is  extended clumpy
  star formation  in major mergers \citep{Powell2013}.  They find that
  most mergers  have an extended  star formation component  during the
  early stages of  a merger, but star formation becomes  nuclear as the
  galaxies  approach  coalescence. \citet{Moreno15}  find that the enhanced star formation in mergers is a combination
of intense enhancements within the central kpc and moderately suppressed activity at larger
galacto-centric radii.  Here  we try  to  distinguish  and
  quantify the  amount of  star formation in  the bulge  (central) and
  disc (extended) components. To do this  we look at the amount of new
  stars formed at the end of the simulation in the remnant compared to
  the  new stars  formed in  the isolation  run. This  additional star
  formation is  then due to  just the merger. Our definition of the SF enhancement include stars formed
in the satellite as well. A more sophisticated analysis of the merger driven star formation rates is presented in a complementary work \citep{Karman2015}, but the general trends are investigated here.
  
\begin{figure}
\centering \includegraphics[scale = 0.45]{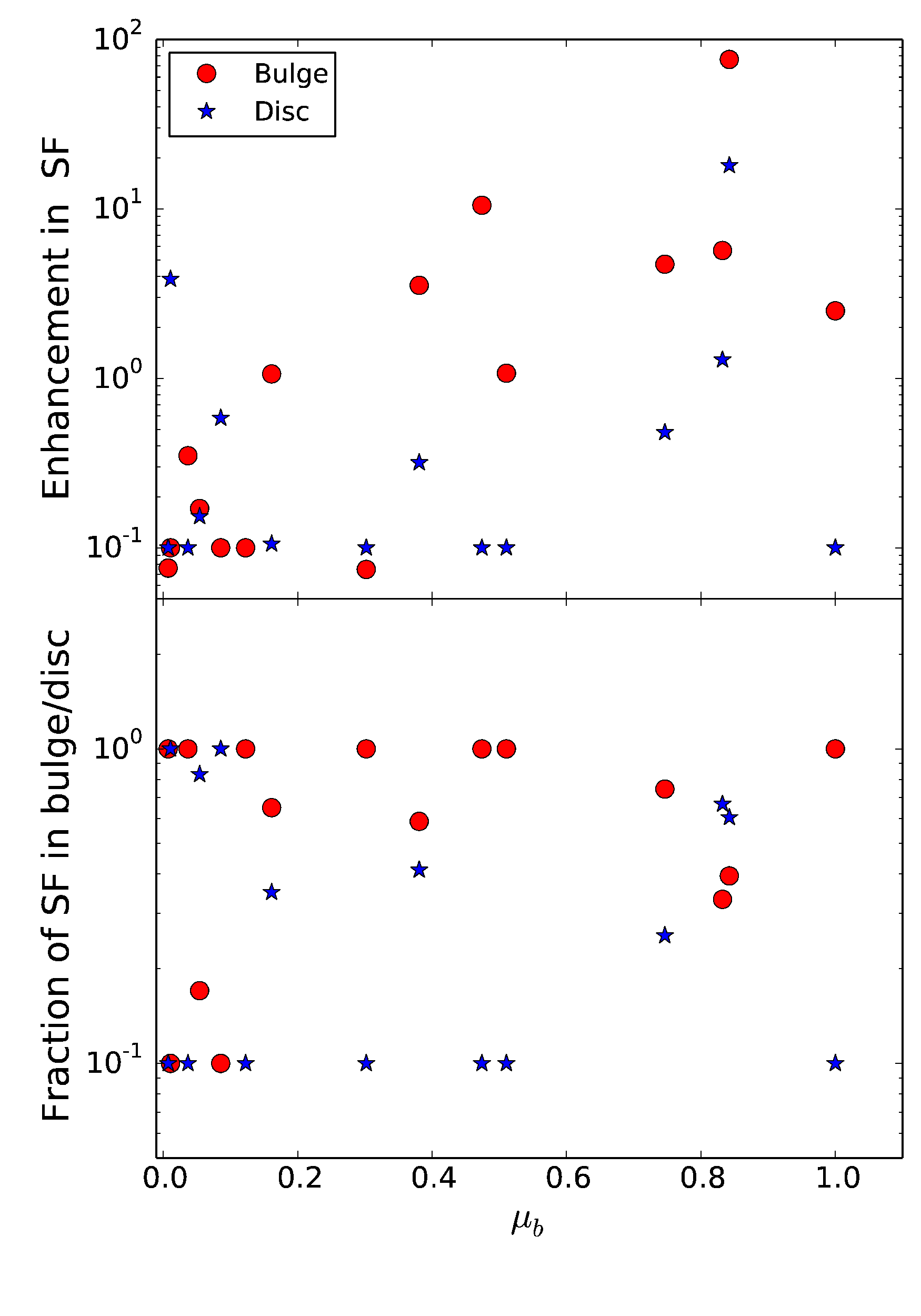}
\caption{Top: The fractional  change in star formation  in the central
  bulge (red circles)  and the central disc (blue  stars). Bottom: The
  fractional distribution of new stars formed due to merger.}
\label{fig:dfbsfr}
\end{figure}

  The top panel  of Fig
\ref{fig:dfbsfr} shows  the fractional change in  star formation rate,
i.e., the star formation in  merger simulation minus star formation in
isolated galaxy divided by the star formation in isolated galaxy, as a
function of  the merger ratio  for both  bulge (red circles)  and disc
(blue stars) component.  We see that in most cases  the star formation
is enhanced both in the bulge and  the disc. For very minor mergers we
see an  increase in  SF by about  $10\%$ in both  the bulge  and disc,
whereas in major mergers  the SF in the bulge can  be enhanced as high
as  $80$  times the  original  value.   On  the  other hand  there  is
relatively low enhancement (of about $20$ times) in the SF in the disc
during a merger.

We  can  then  turn the  previous  plot  around  and  look at,  for  a
particular enhancement in  SF, what fraction of it takes  place in the
bulge and  disc. The  bottom panel  of fig.\ref{fig:dfbsfr}  shows the
fraction of SF enhancement which takes place in the bulge and disc. In
most mergers most of the star formation occurs in the bulge, but quite
a  few  merger scenarios  do  show  that  the  star formation  can  be
triggered in discs as well.
Our results  seem to  suggest that the  relative contribution  of disc
starburst  to the  overall star  formation enhancement  is larger  for
minor mergers. In  order to  confirm and further  quantify this  effect higher
resolution simulations  are needed  that go beyond  the scope  of this
paper.

\subsection{Hot halo}

Previously, simulations which have included hot haloes, have looked into its effects  in both minor mergers and major mergers in non-cosmological setting (\citealt{2011MNRAS.415.3750M}, \citealt{2012MNRAS.423.2045M}) and on the star formation rates \citep{Kim2009}. It  is then interesting  to check what role the hot halo plays in these cosmologically motivated merger simulations.  We  perform a new simulation for Tree  $215240$ (chosen for
its conveniently large merger ratio  $0.77$) without including the hot
halo (both in  the models of the primary and  the satellite galaxies),
but keeping  all other components the  same (stars, cold gas  and dark
matter).

\begin{figure*}
\centering \includegraphics[scale = 0.7]{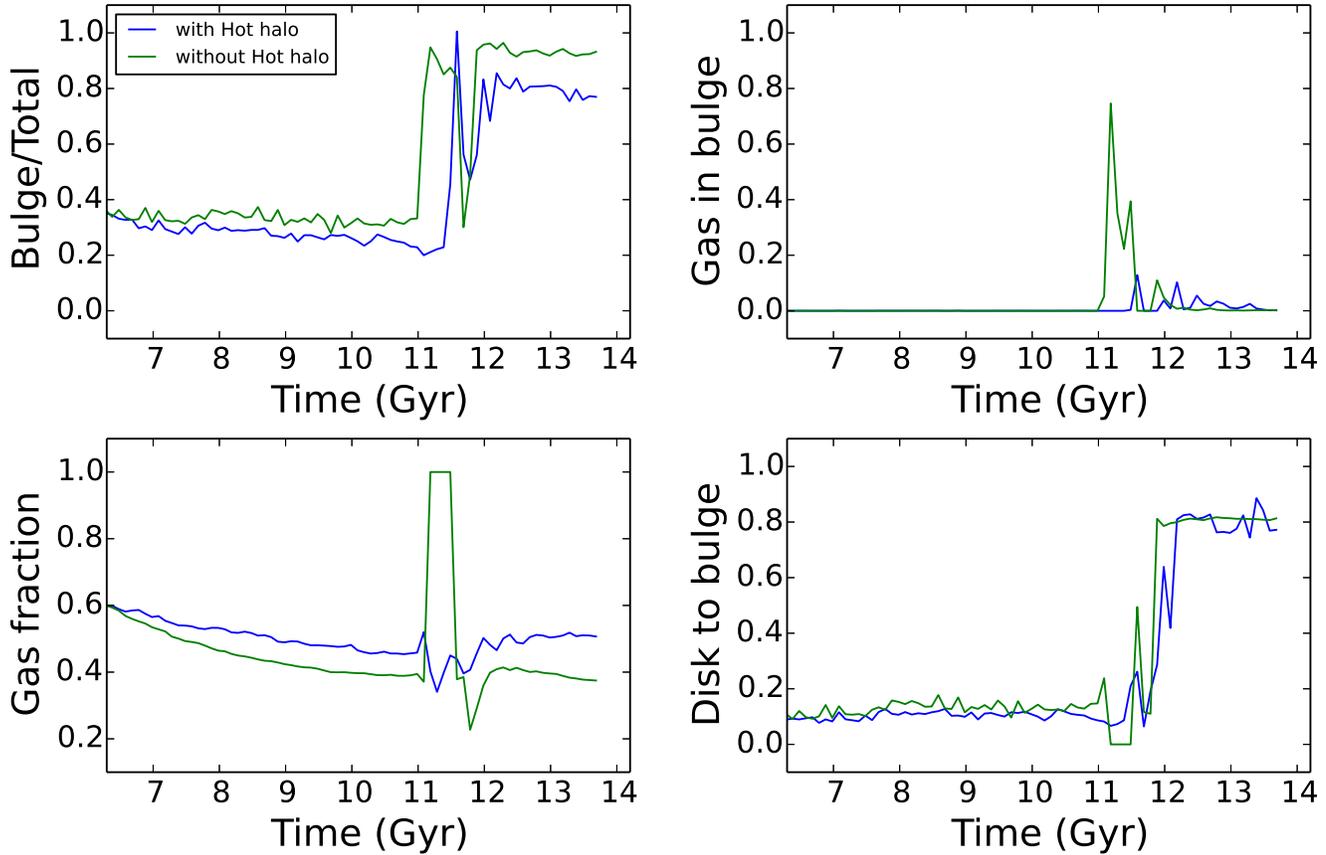}
\caption{Effect of  the hot halo  on the $B/T$, fraction  of satellite
  mass given to central bulge and  fraction of central disc mass given
  to  bulge. The  blue curve  indicates a  simulation including  a hot
  gaseous halo  and the green  curves are simulations without  the hot
  halo.}
\label{fig:nhh}
\end{figure*}

Figure \ref{fig:nhh}  shows the  comparison between  the two  runs for
various properties as  a function of time.  Without a  hot halo (green
curves) the  primary galaxy  becomes completely bulge  dominated after
the  merger,  as shown  in  the  upper  left  panel. The  $B/T$  ratio
increases from a value of about $\approx  0.8$ in the run with the hot
halo (blue  curves) to  $\approx 1.0$  when the  hot gas  reservoir is
removed.  This  is due to the  fact that the hot  halo replenishes the
disc of the  galaxy, keeping it gas  rich as shown on  the bottom left
panel (gas fraction of primary disc).  This in turn helps the gas disc
to  survive  the  merger  (eg.  H09  and  Eq.  \ref{eq:cgas}  of  this
work).   This   can   be   seen    in   the   top   right   panel   of
Fig. \ref{fig:nhh}. We can also see  that the stellar mass transfer to
the bulge from the disc remains  unaffected and does not depend on the
gas  fraction of  the disc  (bottom right  panel). The  hot halo  also
provides  a continuous  supply of  cold gas  for SF  which allows  for
higher      rate      of      disc     regrowth      after      merger
\citep{2011MNRAS.414.1439D}.  All  together  it   is  clear  that  the
inclusion of a hot halo has several  effects on the the final bulge to
total ratio, since  this hot gas reservoir is crucial  for keeping the
disc  gas rich  and  for  disc regrowth  after  a  merger. Hence  this
component is important  and should be taken into  account in numerical
studies      of      mergers      of      galaxies      (see      also
\citealt{2012MNRAS.423.2045M}).

\subsection{Effect of resolution}

\begin{figure}
\centering \includegraphics[scale = 0.45]{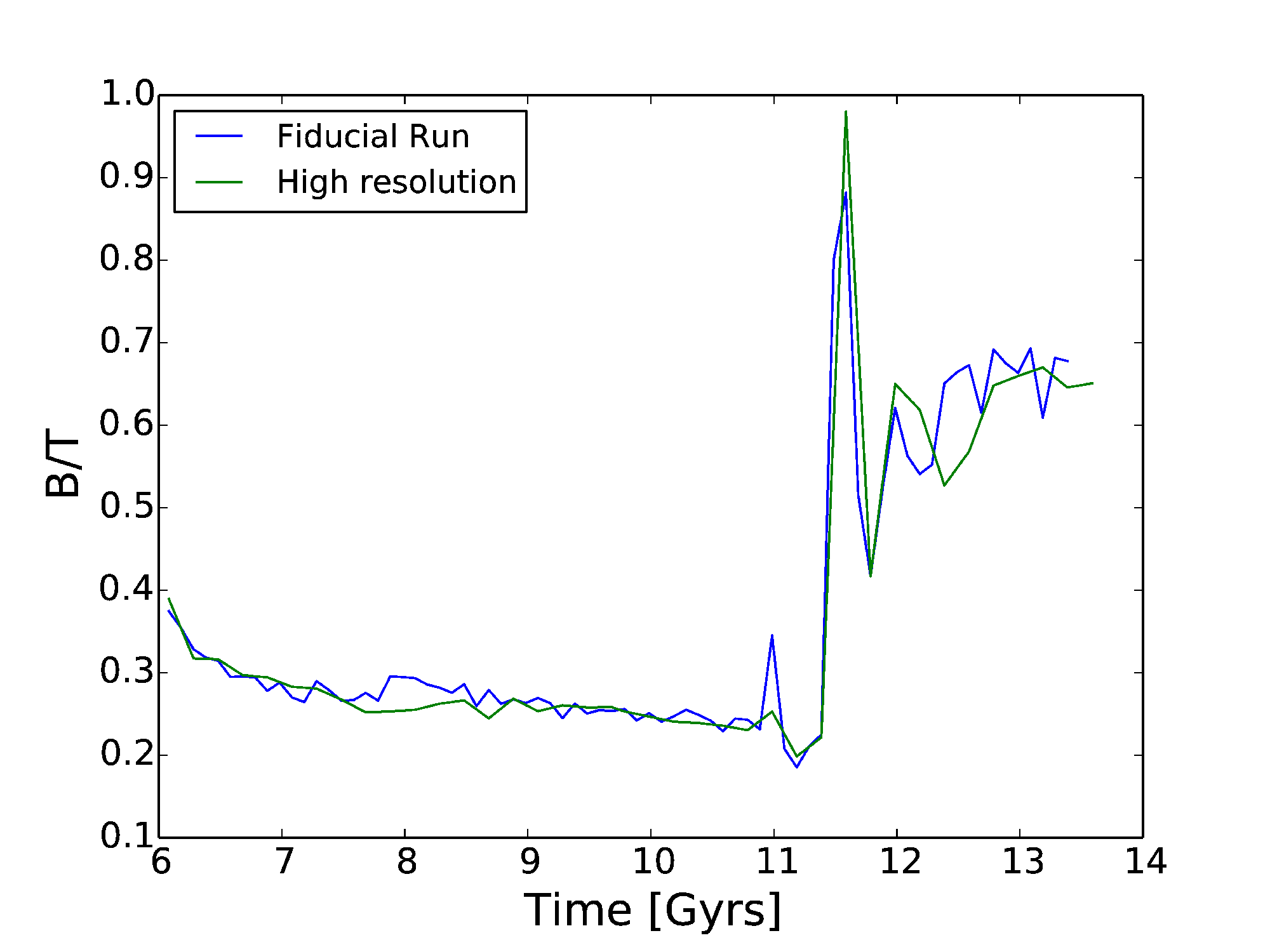}
\caption{The evolution of B/T ratio as a function of time for Tree 215240 ($\mu = 0.88$) in both the low and high resolution run. The high resolution run has two times as many particles as the fiducial run and has $\sim 1.4$ times higher spatial resolution.}
\label{fig:res}
\end{figure}

To test the effect of resolution  we simulated Tree 215240 ($\mu=0.88$) at a higher resolution, containing two times more particles than the fiducial run, i.e., $10^6$ stellar particles in the final merger remnant. This also decreases the softening lengths by a factor of $\sim 1.4$.  All the other simulation parameters are the same as used in the fiducial run. Fig. \ref{fig:res} shows the bulge to total ratio for both the fiducial and high resolution simulations. Encouragingly, the different resolution runs match quite well. There are however, some subtle differences between the runs. The disc of the higher resolution run seems to be a bit more stable to perturbations than the fiducial run. This is evidenced more clearly after the merger takes place.  The disc of the primary in the fiducial run is excited and there seems to be a larger perturbations on the low resolution disc as evidenced by the larger variation of the  B/T ratio on short timescales.  The general behaviour, however,  is quite well converged.  Although, we only show the comparison of the evolution of the B/T ratio, we checked the other mass transfer channels and they all seem to be well converged with respect to resolution as well. 

\section{Discussion and conclusions}
\label{sec:discussion}

Theoretical models  of galaxy  formation and evolution  usually assume
that star formation in galaxies mainly occurs in a disc-like structure
as a consequence  of the conservation of angular  momentum acquired by
their host  DM haloes. In  such a scenario spheroidal  structures like
bulges are created by dissipative  mechanisms, like galaxy mergers and
disc instabilities, which remove angular  momentum from stars and gas,
funnelling them towards the centre of galaxy. Most theoretical models,
like  SAMs, employ  simplified formulae  for dealing  with these  mass
transfer processes,  in the idealized  approximation of a  sequence of
binary mergers.

In  this paper,  we  analyse  in full  detail  the  physics of  galaxy
mergers,  and how  mergers  change the  morphology  of galaxies,  with
particular  emphasis  on  the  mass  transfer  processes  between  the
different  components of  the merging  galaxies.  Our  approach (first
presented and described  in \citealt{2014MNRAS.437.1027M}) consists of
using high  resolution smoothed  particle hydrodynamic  simulations of
galaxy  mergers  where  the  initial   conditions  are  taken  from  a
combination of  cosmological realization of dark  matter haloes merger
trees and  semi-analytic models. The cosmological  merger trees ensure
that  the   orbits  and  merger   timings  of  DM   substructures  are
cosmologically  consistent,   while  the  SAM  modelling   provides  a
reasonable  guess  for the  properties  of  galaxies living  in  these
haloes.

We run  a series of  simulations from  $z=1$ to $z=0$,  including both
single  and   multiple  mergers   and  we   quantify  the   amount  of
morphological transformation in these model galaxies. We decompose the
bulge  and  disc  component  of  model  galaxies  using  all  the  six
dimensional phase space  of position and velocity  and considering the
mass  distribution  as a  function  of  the rotational  support.  This
decomposition  approach is  very robust  and allows  us to  track mass
transfers from  the satellite to the  bulge, from the initial  disc to
the bulge, the mass of the disc dispersed into the stellar halo of the
galaxy and the  $B/T$ evolution. This then allows us  the quantify the
amount of  mass transfer in each  channel as a function  of the merger
parameters.


{\it Mass transfers from the satellite to the bulge:} In major mergers
most  of the  satellite's  mass is  deposited into  the  bulge of  the
central galaxy. On  the other hand, a small  satellite galaxy deposits
most of its baryonic mass into  the halo (due to stellar stripping and
ram pressure) before its final coalescence with the central object.

{\it Mass transfers  from the central disc to the  central bulge:} The
interaction with the incoming satellite causes matter inflows from the
central disc  to the central  bulge. We  confirm the findings  of H09,
that   a   gas  rich   disc   is   able   to  survive   mergers   more
effectively.   We note here that we do not resolve the small scale physics of the ISM, instead choosing to model it using an effective EOS \citep{2003MNRAS.339..289S}. This model provides a high thermal pressure support in the ISM, which smoothes out any instabilities in the disc. This model works quite well for low redshift redshift galaxies which are not highly star forming. At high redshifts  the ISM is usually very turbulent and clumpy.  Recent studies have shown that masses and sizes of disc components that survive a major merger are severely reduced when the hydrodynamics of the high redshift ISM are modelled properly \citep{Bournaud2011}. 
Since the starting point of our simulations is $z=1$, the disc is expected to be much more calm. Therefore the ISM model used in our work is quite sufficient to capture the physics of mergers at the redshifts we consider.  

However, our  results show  that the  amount of  central
stellar disc mass transferred to the central bulge is quite low.  Even
for $1:1$ mergers only $37\%$ of  the central stellar disc is given to
the central bulge.  In addition a fraction of the central disc mass is
expelled outwards due to  close gravitational encounters (about $22\%$
for  $1:1$   mergers),  adding   up  to   the  diffuse   stellar  halo
population. This mass transfer channel is generally poorly modelled in
many SAMs.

{\it Merger  driven star  formation: } Our  results also  confirm that
there is a  sharp enhancement of star formation during  a merger. Many
SAMs assume  that all of the  merger driven starburst takes  places in
the bulge.  However, our
simulations  echo recent results \citep{Powell2013,Moreno15}, which show that there  is an  enhancement of  star formation  in the
discs  of central  galaxies  in  addition to  the  enhancement in  the
bulge. As expected, larger starbursts are attributed to major mergers,
but the  relative contribution of  disc starburst to the  overall star
formation  enhancement  is  larger  for   minor  mergers.

{\it Effect of hot halo: }One  of the major improvements of this work,
with respect to previous similar attempts,  lies in the modelling of a
hot  gaseous halo  component: the  existence of  such a  component (in
addition  to the  cold phase  present in  the disc  of the  galaxy) is
predicted            by            cosmological            simulations
(e.g.   \citealt{2006ApJ...644L...1S};  \citealt{2009ApJ...697L..38J};
\citealt{2009MNRAS.399..239R};    \citealt{2011ApJ...734...62H})   and
observations            (e.g.           \citealt{2009MNRAS.394.1741O},
\citealt{2013ApJ...762..106A}).   The  hot  gaseous halo  affects  the
final  bulge to  total ratio  of the  galaxy: in  fact it  is able  to
replenish the  galaxy disc with  fresh infalling cold gas,  keeping it
gas rich and fuelling disc star formation over longer time scales.

We thus propose  a series of new fitting formulae  able to capture the
trends in  the relations between  the amplitude of mass  transfers and
baryonic (stars  + cold gas)  merger ratio. We  are well aware  of the
limitation of  our analysis,  a relatively  small sample  of simulated
haloes.   More work  is of  course needed  to increase  the simulation
sample,  including more  merger  trees at  different  mass scales  and
widening  the range  of orbital/merger  parameter. This  larger sample
would  then  provide  us  with  better constrains  on  both  the  mean
relations and their  scatter, as well as a testbed  for extreme cases,
such as multiple merger scenarios. Despite the small sample considered
in this  paper, our results are  indicative of a need  for revision in
our understanding  of mass  flows involved  in a  galaxy merger.  In a
work (Fontanot et al., in preparation), we plan to include
our  fitting   formulae  in   state-of-the-art  SAMs,  to   study  the
implications of  our findings  on a cosmologically  significant galaxy
sample,  in  determining the  fraction  of  bulge and  disc  dominated
galaxies.

Overall  our  study  shows  that the combined effect of satellite accretion and disc destruction in building bulges is not as  efficient  as
previously thought and this will
possibly alleviate some of the  tensions between the observed fraction
of  bulgeless galaxies  and  the hierarchical  scenario for  structure
formation as predicted by the $\Lambda$CDM model.

\section*{Acknowledgments} 

The numerical simulations used in this work were performed on the THEO
cluster   of  the   Max-Planck-Institut   f\"ur   Astronomie  at   the
Rechenzentrum in Garching.  RK and  AVM acknowledge the support of SFB
881  (subproject  A1) of  the  German  research foundation  (DFG).  FF
acknowledges financial  support from the Klaus  Tschira Foundation and
the Deutsche Forschungsgemeinschaft through  Transregio 33, ``The Dark
Universe''.


\bibliographystyle{mn2e} \bibliography{kannan14}

\label{lastpage}

\end{document}